\documentclass[prb,twocolumn,superscriptaddress,floatfix,letterpaper]{revtex4-1}
\usepackage[all,cmtip]{xy}
\newcommand{\x}{\underline}

\newcommand{\rep}{\mathrm{Rep}}
\newcommand{\vect}{\mathrm{Vec}}
\newcommand{\Cb}{\mathbb{C}}

\graphicspath{{figs/}}
\newcommand{\frc}[2]{{{#1}/{#2}}}

\newcommand\void[1]       {}
\newcommand{\cen}[2]{#1^\text{cen}_{#2}}
\newcommand{\repgz}[2]{\mathrm{sRep}(#1#2)}

\theoremstyle{wenthm}

\newtheorem{rema}{Remark}
\newtheorem{expl}{Example}

\begin{document}

\title{A theory of 2+1D fermionic topological orders\\
and fermionic/bosonic topological orders with symmetries
}

\author{Tian Lan} 

\affiliation{Perimeter Institute for Theoretical Physics, Waterloo, Ontario N2L
2Y5, Canada} 
\affiliation{Department of Physics and Astronomy,
  University of Waterloo, Waterloo, Ontario N2L 3G1, Canada}

\author{Liang Kong} 
\affiliation{Department of Mathematics \& Statistics, University of New Hampshire, Durham, NH, 03824, USA}
\affiliation{Center of Mathematical Sciences and Applications, 
Harvard University, Cambridge, MA 02138}

\author{Xiao-Gang Wen} 

\affiliation{Department of Physics, Massachusetts Institute of Technology, Cambridge, Massachusetts 02139, USA}
\affiliation{Perimeter Institute for Theoretical Physics, Waterloo, Ontario N2L
2Y5, Canada}

\begin{abstract} 
We propose that, up to invertible topological orders, 2+1D fermionic
topological orders without symmetry and 2+1D fermionic/bosonic topological
orders with symmetry $G$ are classified by non-degenerate unitary braided
fusion categories (UBFC) over a symmetric fusion category (SFC); the SFC
describes a fermionic product state without symmetry or a fermionic/bosonic
product state with symmetry $G$, and the UBFC has a modular extension.
We developed a simplified theory of non-degenerate UBFC over a SFC based on the
fusion coefficients $N^{ij}_k$ and spins $s_i$.  This allows us to obtain a
list that contains all 2+1D fermionic topological orders (without symmetry).
We find explicit realizations for all the fermionic topological orders in the
table.  For example, we find that, up to invertible $p+\ii p$ fermionic
topological orders, there are only four fermionic topological orders with one
non-trivial topological excitation: (1) the $K={\scriptsize \bpm
-1&0\\0&2\epm}$ fractional quantum Hall state, (2) a Fibonacci bosonic
topological order $2^B_{14/5}$ stacking with a fermionic product state, (3) the
time-reversal conjugate of the previous one, (4) a primitive fermionic
topological order that has a chiral central charge $c=\frac14$, whose only
topological excitation has a non-abelian statistics with a spin $s=\frac14$ and
a quantum dimension $d=1+\sqrt{2}$.  We also proposed a categorical way to
classify 2+1D invertible fermionic topological orders using modular extensions.

\end{abstract}

\maketitle

{\small \setcounter{tocdepth}{2} \tableofcontents }

\section{Introduction}

\subsection{Background}

Topological order\cite{Wtop,WNtop,Wrig} is a new kind of order beyond Landau
symmetry breaking theory. It cannot be characterized by the local order
parameters associated with the symmetry breaking.  However, topological order
can be characterized/defined by (a) the topology-dependent ground state
degeneracy\cite{Wtop,WNtop} and (b) the non-abelian geometric phases $(S,T)$ of
the degenerate ground states\cite{Wrig,KW9327}. Those quantities are
\emph{robust against any local perturbations}.\cite{WNtop} Thus they are
topological invariants that define new kind of quantum phases -- topologically
ordered phases.  Recently, it was found that, microscopically, topological
order is related to long-range entanglement.\cite{LW0605,KP0604} In fact, we
can regard topological orders as patterns of long-range entanglement in
many-body ground states,\cite{CGW1038} which is defined as the equivalence
classes of gapped  quantum liquid\cite{ZW1490} states under local unitary
transformations.\cite{LWstrnet,VCL0501,V0705} Chiral spin
liquids\cite{KL8795,WWZ8913}, integral/fractional quantum Hall
states\cite{KDP8094,TSG8259,L8395}, $\Z_2$ spin
liquids\cite{RS9173,W9164,MS0181}, non-abelian fractional quantum Hall
states\cite{MR9162,W9102,WES8776,RMM0899}, etc., are examples of topologically
ordered phases.

Topological order and long-range entanglement are truly new phenomena, which
require new mathematical language to describe them.  Tensor category
theory\cite{FNS0428,LWstrnet,CGW1038,GWW1017,KK1251,GWW1332} and simple current
algebras\cite{MR9162,BW9215,WW9455,LWW1024} (or patterns of zeros
\cite{WW0808,WW0809,BW0932,SL0604,BKW0608,SY0802,BH0802,BH0802a,BH0882}) may be
parts of the new mathematical language.  Using the new mathematical language,
some systematic classification results for certain type of topological orders
in low dimensions were achieved.

Using unitary fusion category (UFC) theory, we have developed a systematic and
quantitative theory that classifies all the topological orders with gappable
edge for 2+1D interacting bosonic systems.\cite{LWstrnet,CGW1038} A double
Fibonacci bosonic topological order $2^B_{14/5}\boxtimes 2^B_{-14/5}$ was
discovered.\cite{LWstrnet} We also developed a fermionic UFC theory, to
classify topological orders with gappable edge for 2+1D interacting fermionic
systems.\cite{GWW1017,GWW1332} For 2+1D bosonic/fermionic topological orders
(with gappable or un-gappable edge) that have only abelian statistics, we find
that we can use integer $K$-matrices to classify them and use the following
$U(1)$ Chern-Simons theory to describe
them\cite{BW9045,R9002,FK9169,WZ9290,KW9327,BM0535,KS1193}
\begin{align}
\label{csK}
 {\cal L}= \frac{K_{IJ}}{4\pi} a_{I\mu} \prt_\nu a_{J\la}\eps^{\mu\nu\la} .
\end{align}
Such an effective theory can be realized by multi-layer fractional quantum Hall
state:
\begin{align}
\prod_{I;i<j} (z_i^I-z_j^I)^{K_{II}}
\prod_{I<J;i,j} (z_i^I-z_j^J)^{K_{IJ}}
\ee^{-\frac14 \sum_{i,I} |z_i^I|^2}.
\end{align}
When diagonal $K_{II}$'s are all even, the $K$-matrices classify 2+1D bosonic
abelian topological orders.  When some diagonal $K_{II}$'s are odd, the
$K$-matrices classify 2+1D fermionic abelian topological orders.

\def\arraystretch{1.22} \setlength\tabcolsep{3pt}
\begin{table*}[tbp] 
\caption{
A list of simple fermionic topological orders (up to invertible ones) with  $N$
types of topological excitations (including the parent fermion) and  chiral
central charge $c$ (mod 1/2). The excitations have quantum dimension $d_i$ and
spin $s_i$ (mod 1).  The searched maximum $\t N^{[i][j]}_{[k]}$ is 8 for $N=4$,
5 for $N=6$, an 4 for $N=8$.  Here
$\zeta_n^m=\frac{\sin[\pi(m+1)/(n+2)]}{\sin[\pi/(n+2)]}$.  
} 
\label{toplst} 
\centering
\begin{tabular}{ |c|c|c|l|l|c| } 
\hline 
$N^F_c$ & $S_\text{top}$ & $D^2$ & $d_1,d_2,\cdots$ & $s_1,s_2,\cdots$ & comments \\
 \hline 
$2^F_{ 0}$ & $0$ & $2$ & $1,1$ & $0, \frac{1}{2}$ & trivial $\cF_0$\\
 \hline 
$4^F_{ 0}$ & $0.5$ & $4$ & $1,1,1,1$ & $0, \frac{1}{2}, \frac{1}{4},-\frac{1}{4}$ & $\cF_0\boxtimes 2^B_{ 1}$\\
$4^F_{\frc 15}$ & $0.9276$ & $7.2360$ & $1,1,\zeta_{3}^{1},\zeta_{3}^{1}$ & $0, \frac{1}{2}, \frac{1}{10},-\frac{2}{5}$ & $\cF_0\boxtimes 2^B_{-\frc{14}{5}}$\\
$4^F_{-\frc 15}$ & $0.9276$ & $7.2360$ & $1,1,\zeta_{3}^{1},\zeta_{3}^{1}$ & $0, \frac{1}{2},-\frac{1}{10}, \frac{2}{5}$ & $\cF_0\boxtimes 2^B_{\frc{14}{5}}$\\
$4^F_{1/4}$ & $1.3857$ & $13.6568$ & $1,1,\zeta_{6}^{2},\zeta_{6}^{2}$ & $0, \frac{1}{2},-\frac{1}{4}, \frac{1}{4}$ & $\cF_{(A_1,6)}$\\
\red{$4^F_{ *}$} & \red{$2.1218$} & \red{$37.888$} & \red{$1,1,2+\sqrt{5},2+\sqrt{5}$} & \red{$0, \frac{1}{2}, \frac{1}{4},-\frac{1}{4}$} & \red{invalid}\\
\red{$4^F_{ *}$} & \red{$2.6422$} & \red{$77.946$} & \red{$1,1,3+\sqrt{10},3+\sqrt{10}$} & \red{$0, \frac{1}{2},-\frac{1}{4}, \frac{1}{4}$} & \red{invalid}\\
\red{$4^F_{ *}$} & \red{$3.0328$} & \red{$133.968$} & \red{$1,1,4+\sqrt{17},4+\sqrt{17}$} & \red{$0, \frac{1}{2}, \frac{1}{4},-\frac{1}{4}$} & \red{invalid}\\
 \hline 
$6^F_{ 0}$ & $0.7924$ & $6$ & $1,1,1,1,1,1$ & $0, \frac{1}{2}, \frac{1}{6},-\frac{1}{3}, \frac{1}{6},-\frac{1}{3}$ & $\cF_0\boxtimes 3^B_{-2}$\\
$6^F_{ 0}$ & $0.7924$ & $6$ & $1,1,1,1,1,1$ & $0, \frac{1}{2},-\frac{1}{6}, \frac{1}{3},-\frac{1}{6}, \frac{1}{3}$ & $\cF_0\boxtimes 3^B_{ 2}$\\
$6^F_{ 0}$ & $1$ & $8$ & $1,1,1,1,\zeta_{2}^{1},\zeta_{2}^{1}$ & $0, \frac{1}{2}, 0, \frac{1}{2}, \frac{1}{16},-\frac{7}{16}$ & $\cF_0\boxtimes 3^B_{\frc{1}{2}}$\\
$6^F_{ 0}$ & $1$ & $8$ & $1,1,1,1,\zeta_{2}^{1},\zeta_{2}^{1}$ & $0, \frac{1}{2}, 0, \frac{1}{2}, \frac{3}{16},-\frac{5}{16}$ & $\cF_0\boxtimes 3^B_{\frc{3}{2}}$\\
$6^F_{ 0}$ & $1$ & $8$ & $1,1,1,1,\zeta_{2}^{1},\zeta_{2}^{1}$ & $0, \frac{1}{2}, 0, \frac{1}{2}, \frac{5}{16},-\frac{3}{16}$ & $\cF_0\boxtimes 3^B_{-\frc{3}{2}}$\\
$6^F_{ 0}$ & $1$ & $8$ & $1,1,1,1,\zeta_{2}^{1},\zeta_{2}^{1}$ & $0, \frac{1}{2}, 0, \frac{1}{2}, \frac{7}{16},-\frac{1}{16}$ & $\cF_0\boxtimes 3^B_{-\frc{1}{2}}$\\
$6^F_{\frc17}$ & $1.6082$ & $18.5916$ & $1,1,\zeta_{5}^{1},\zeta_{5}^{1},\zeta_{5}^{2},\zeta_{5}^{2}$ & $0, \frac{1}{2}, \frac{5}{14},-\frac{1}{7},-\frac{3}{14}, \frac{2}{7}$ & $\cF_0\boxtimes 3^B_{\frc{8}{7}}$\\
$6^F_{-\frc17}$ & $1.6082$ & $18.5916$ & $1,1,\zeta_{5}^{1},\zeta_{5}^{1},\zeta_{5}^{2},\zeta_{5}^{2}$ & $0, \frac{1}{2},-\frac{5}{14}, \frac{1}{7}, \frac{3}{14},-\frac{2}{7}$ & $\cF_0\boxtimes 3^B_{-\frc{8}{7}}$\\
$6^F_{ 0}$ & $2.2424$ & $44.784$ & $1,1,\zeta_{10}^{2},\zeta_{10}^{2},\zeta_{10}^{4},\zeta_{10}^{4}$ & $0, \frac{1}{2}, \frac{1}{3},-\frac{1}{6}, 0, \frac{1}{2}$ & primitive\\
$6^F_{ 0}$ & $2.2424$ & $44.784$ & $1,1,\zeta_{10}^{2},\zeta_{10}^{2},\zeta_{10}^{4},\zeta_{10}^{4}$ & $0, \frac{1}{2},-\frac{1}{3}, \frac{1}{6}, 0, \frac{1}{2}$ & $\cF_{(A_1,10)}$\\
 \hline 
$8^F_{ 0}$ & $1$ & $8$ & $1,1,1,1,1,1,1,1$ & $0, \frac{1}{2}, 0, \frac{1}{2}, \frac{1}{8},-\frac{3}{8}, \frac{1}{8},-\frac{3}{8}$ & $\cF_0\boxtimes 4^B_{ 1}$\\
$8^F_{ 0}$ & $1$ & $8$ & $1,1,1,1,1,1,1,1$ & $0, \frac{1}{2}, 0, \frac{1}{2},-\frac{1}{4}, \frac{1}{4},-\frac{1}{4}, \frac{1}{4}$ & $\cF_0\boxtimes 4^B_{ 0}$\\
$8^F_{ 0}$ & $1$ & $8$ & $1,1,1,1,1,1,1,1$ & $0, \frac{1}{2}, 0, \frac{1}{2}, \frac{3}{8},-\frac{1}{8}, \frac{3}{8},-\frac{1}{8}$ & $\cF_0\boxtimes 4^B_{-1}$\\
$8^F_{ 0}$ & $1$ & $8$ & $1,1,1,1,1,1,1,1$ & $0, \frac{1}{2}, 0, \frac{1}{2}, \frac{1}{2}, 0, \frac{1}{2}, 0$ & $\cF_0\boxtimes 4^B_{ 0}$\\
$8^F_{\frc15}$ & $1.4276$ & $14.4720$ & $1,1,1,1,\zeta_{3}^{1},\zeta_{3}^{1},\zeta_{3}^{1},\zeta_{3}^{1}$ & $0, \frac{1}{2},-\frac{1}{4}, \frac{1}{4}, \frac{1}{10},-\frac{2}{5}, \frac{7}{20},-\frac{3}{20}$ & $\cF_0\boxtimes 4^B_{-\frc{9}{5}}$\\
$8^F_{-\frc15}$ & $1.4276$ & $14.4720$ & $1,1,1,1,\zeta_{3}^{1},\zeta_{3}^{1},\zeta_{3}^{1},\zeta_{3}^{1}$ & $0, \frac{1}{2},-\frac{1}{4}, \frac{1}{4},-\frac{1}{10}, \frac{2}{5}, \frac{3}{20},-\frac{7}{20}$ & $\cF_0\boxtimes 4^B_{\frc{9}{5}}$\\
$8^F_{ 0}$ & $1.7924$ & $24$ & $1,1,1,1,2,2,\sqrt{6},\sqrt{6}$ & $0, \frac{1}{2}, \frac{1}{2}, 0, \frac{1}{6},-\frac{1}{3},-\frac{1}{16}, \frac{7}{16}$ & primitive\\
$8^F_{ 0}$ & $1.7924$ & $24$ & $1,1,1,1,2,2,\sqrt{6},\sqrt{6}$ & $0, \frac{1}{2}, \frac{1}{2}, 0, \frac{1}{6},-\frac{1}{3}, \frac{5}{16},-\frac{3}{16}$ & primitive\\
$8^F_{ 0}$ & $1.7924$ & $24$ & $1,1,1,1,2,2,\sqrt{6},\sqrt{6}$ & $0, \frac{1}{2}, \frac{1}{2}, 0, \frac{1}{6},-\frac{1}{3},-\frac{5}{16}, \frac{3}{16}$ & primitive\\
$8^F_{ 0}$ & $1.7924$ & $24$ & $1,1,1,1,2,2,\sqrt{6},\sqrt{6}$ & $0,
\frac{1}{2}, \frac{1}{2}, 0, \frac{1}{6},-\frac{1}{3},-\frac{7}{16},
\frac{1}{16}$ & $\cF_{U(1)_6/\Z_2}$\\
$8^F_{ 0}$ & $1.7924$ & $24$ & $1,1,1,1,2,2,\sqrt{6},\sqrt{6}$ & $0, \frac{1}{2}, \frac{1}{2}, 0,-\frac{1}{6}, \frac{1}{3},-\frac{1}{16}, \frac{7}{16}$ & primitive\\
$8^F_{ 0}$ & $1.7924$ & $24$ & $1,1,1,1,2,2,\sqrt{6},\sqrt{6}$ & $0, \frac{1}{2}, \frac{1}{2}, 0,-\frac{1}{6}, \frac{1}{3}, \frac{3}{16},-\frac{5}{16}$ & primitive\\
$8^F_{ 0}$ & $1.7924$ & $24$ & $1,1,1,1,2,2,\sqrt{6},\sqrt{6}$ & $0, \frac{1}{2}, \frac{1}{2}, 0,-\frac{1}{6}, \frac{1}{3},-\frac{3}{16}, \frac{5}{16}$ & primitive\\
$8^F_{ 0}$ & $1.7924$ & $24$ & $1,1,1,1,2,2,\sqrt{6},\sqrt{6}$ & $0, \frac{1}{2}, \frac{1}{2}, 0,-\frac{1}{6}, \frac{1}{3},-\frac{7}{16}, \frac{1}{16}$ & primitive\\
$8^F_{-1/10}$ & $1.8552$ & $26.180$ & $1,1,\zeta_{3}^{1},\zeta_{3}^{1},\zeta_{3}^{1},\zeta_{3}^{1},\zeta_{8}^{2},\zeta_{8}^{2}$ & $0, \frac{1}{2}, \frac{1}{10},-\frac{2}{5}, \frac{1}{10},-\frac{2}{5},-\frac{3}{10}, \frac{1}{5}$ & $\cF_0\boxtimes 4^B_{\frc{12}{5}}$\\
$8^F_{0}$ & $1.8552$ & $26.180$ & $1,1,\zeta_{3}^{1},\zeta_{3}^{1},\zeta_{3}^{1},\zeta_{3}^{1},\zeta_{8}^{2},\zeta_{8}^{2}$ & $0, \frac{1}{2}, \frac{1}{10},-\frac{2}{5},-\frac{1}{10}, \frac{2}{5}, \frac{1}{2}, 0$ & $\cF_0\boxtimes 4^B_{ 0}$\\
$8^F_{1/10}$ & $1.8552$ & $26.180$ & $1,1,\zeta_{3}^{1},\zeta_{3}^{1},\zeta_{3}^{1},\zeta_{3}^{1},\zeta_{8}^{2},\zeta_{8}^{2}$ & $0, \frac{1}{2},-\frac{1}{10}, \frac{2}{5},-\frac{1}{10}, \frac{2}{5}, \frac{3}{10},-\frac{1}{5}$ & $\cF_0\boxtimes 4^B_{-\frc{12}{5}}$\\
$8^F_{1/4}$ & $1.8857$ & $27.312$ & $1,1,1,1,\zeta_{6}^{2},\zeta_{6}^{2},\zeta_{6}^{2},\zeta_{6}^{2}$ & $0, \frac{1}{2},-\frac{1}{4}, \frac{1}{4}, \frac{1}{4},-\frac{1}{4}, \frac{1}{2}, 0$ & $4^F_{1/4}\boxtimes 2^B_1$ \\
$8^F_{1/6}$ & $2.1328$ & $38.468$ & $1,1,\zeta_{7}^{1},\zeta_{7}^{1},\zeta_{7}^{2},\zeta_{7}^{2},\zeta_{7}^{3},\zeta_{7}^{3}$ & $0, \frac{1}{2}, \frac{1}{6},-\frac{1}{3}, \frac{5}{18},-\frac{2}{9},-\frac{1}{6}, \frac{1}{3}$ & $\cF_0\boxtimes 4^B_{-\frc{10}{3}}$\\
$8^F_{-1/6}$ & $2.1328$ & $38.468$ & $1,1,\zeta_{7}^{1},\zeta_{7}^{1},\zeta_{7}^{2},\zeta_{7}^{2},\zeta_{7}^{3},\zeta_{7}^{3}$ & $0, \frac{1}{2},-\frac{1}{6}, \frac{1}{3},-\frac{5}{18}, \frac{2}{9}, \frac{1}{6},-\frac{1}{3}$ & $\cF_0\boxtimes 4^B_{\frc{10}{3}}$\\
$8^F_{-1/20}$ & $2.3133$ & $49.410$ & $1,1,\zeta_{3}^{1},\zeta_{3}^{1},\zeta_{6}^{2},\zeta_{6}^{2},\zeta_3^1\zeta_6^2,\zeta_3^1\zeta_6^2$ & $0, \frac{1}{2}, \frac{1}{10},-\frac{2}{5},-\frac{1}{4}, \frac{1}{4}, \frac{7}{20},-\frac{3}{20}$ & $4^F_{1/4}\boxtimes_{\cF_0} 4^F_{1/5}$\\
$8^F_{1/20}$ & $2.3133$ & $49.410$ & $1,1,\zeta_{3}^{1},\zeta_{3}^{1},\zeta_{6}^{2},\zeta_{6}^{2},\zeta_3^1\zeta_6^2,\zeta_3^1\zeta_6^2$ & $0, \frac{1}{2},-\frac{1}{10}, \frac{2}{5},-\frac{1}{4}, \frac{1}{4}, \frac{3}{20},-\frac{7}{20}$ & $4^F_{1/4}\boxtimes_{\cF_0} 4^F_{-1/5}$\\
\color{red} $8^F_{ *}$ &\color{red}  $2.6218$ &\color{red}  $75.777$ &\color{red}  $1,1,1,1, 2+\sqrt 5, 2+\sqrt 5, 2+\sqrt 5, 2+\sqrt 5$ &\color{red}  $0, \frac{1}{2}, \frac{1}{4},-\frac{1}{4}, 0, \frac{1}{2}, \frac{1}{4},-\frac{1}{4}$ & \color{red} invalid\\
$8^F_{ 0}$ & $2.7715$ & $93.254$ & $1,1,\zeta_{6}^{2},\zeta_{6}^{2},\zeta_{6}^{2},\zeta_{6}^{2},\zeta_6^2\zeta_6^2,\zeta_6^2\zeta_6^2$ & $0, \frac{1}{2}, \frac{1}{4},-\frac{1}{4}, \frac{1}{4},-\frac{1}{4}, 0, \frac{1}{2}$ & $4^F_{1/4}\boxtimes_{\cF_0} 4^F_{1/4}$ \\
$8^F_{-1/8}$ & $2.8577$ & $105.096$ & $1,1,\zeta_{14}^{2},\zeta_{14}^{2},\zeta_{14}^{4},\zeta_{14}^{4},\zeta_{14}^{6},\zeta_{14}^{6}$ & $0, \frac{1}{2}, \frac{3}{8},-\frac{1}{8}, \frac{1}{8},-\frac{3}{8},-\frac{1}{4}, \frac{1}{4}$ & primitive\\
$8^F_{1/8}$ & $2.8577$ & $105.096$ & $1,1,\zeta_{14}^{2},\zeta_{14}^{2},\zeta_{14}^{4},\zeta_{14}^{4},\zeta_{14}^{6},\zeta_{14}^{6}$ & $0, \frac{1}{2},-\frac{3}{8}, \frac{1}{8},-\frac{1}{8}, \frac{3}{8}, \frac{1}{4},-\frac{1}{4}$ & $\cF_{(A_1,14)}$\\
\color{red}$8^F_{ *}$ &\color{red}  $3.0494$ &\color{red}  $137.08$ &\color{red}  {\scriptsize $1,1,\zeta_{3}^{1},\zeta_{3}^{1}, 2+\sqrt 5, 2+\sqrt 5,\zeta_3^1(2+\sqrt 5),\zeta_3^1(2+\sqrt 5)$} &\color{red}  $0, \frac{1}{2}, \frac{1}{10},-\frac{2}{5}, \frac{1}{4},-\frac{1}{4},-\frac{3}{20}, \frac{7}{20}$ &\color{red} invalid \\
\color{red}$8^F_{ *}$ &\color{red}  $3.0494$ &\color{red}  $137.08$ &\color{red} {\scriptsize $1,1,\zeta_{3}^{1},\zeta_{3}^{1}, 2+\sqrt 5, 2+\sqrt 5,\zeta_3^1(2+\sqrt 5),\zeta_3^1(2+\sqrt 5)$} &\color{red}  $0, \frac{1}{2},-\frac{1}{10}, \frac{2}{5}, \frac{1}{4},-\frac{1}{4},-\frac{7}{20}, \frac{3}{20}$ &\color{red} invalid \\
 \hline 
\end{tabular} 
\end{table*}

\subsection{Invertible topological orders}

We can stack two topologically ordered states together to form a new
topologically ordered state.  Such a stacking operation $\boxtimes$ makes the
set of various topological orders into a commutative monoid.\cite{KW1458}  (A
monoid is almost a group except that elements may not have inverses.)
A state has a trivial topological order if the stacking of such state with any
other topological order give the same topological order back.  It turns out that
the states with a trivial topological order are always product states or
short-range entangled states.

Although most topological orders do not have an inverse with respect to the
stacking operation, some topological orders can have an inverse.  Those
topological orders are called invertible topological orders.\cite{KW1458,F1478}
(A topological order $\cC$ is invertible if there exists another topological
order $\cD$, such that the stacking of $\cC$ and $\cD$ gives rise to a trivial
topological order $\one$, \ie $\cC\boxtimes \cD = \one$.)  In fact, such an
inverse $\cD$ can be obtained from $\cC$ by a time-reversal transformation.

It turns out that a topological order is invertible iff it has no non-trivial
topological excitations.\cite{KW1458,F1478} In 2+1D, the set of all invertible
bosonic topological orders form an abelian group $\Z$, which
is generated, via the stacking and time-reversal operations, by the $E_8$
bosonic quantum Hall state described by the following $K$-matrix:
\begin{align}
\label{E8}
K_{E_8}={\footnotesize \begin{pmatrix}
2&1&0&0&0&0&0&0\\ 
1&2&1&0&0&0&0&0\\ 
0&1&2&1&0&0&0&0\\ 
0&0&1&2&1&0&0&0\\
0&0&0&1&2&1&0&1\\ 
0&0&0&0&1&2&1&0\\ 
0&0&0&0&0&1&2&0\\ 
0&0&0&0&1&0&0&2\\
\end{pmatrix} }  .
\end{align}
The $E_8$ bosonic quantum Hall state has no non-trivial topological excitations
(since det$(K)=1$). But the state has a non-trivial thermal Hall
effect\cite{KF9732} and ungappable gapless chiral edge
states\cite{Wedge,Wtoprev} with a chiral central charge $c=8$.  Thus the $E_8$
state has a non-trivial invertible topological order.

\subsection{Classify topological orders via (non-)abelian statistics}

If we overlook the invertible topological orders, \ie consider only the
quotient 
\begin{align}
\frac{ \text{topological orders} }{ \text{ invertible topological orders} }, 
\end{align}
then we can use (non-)abelian statistics of topological excitations to describe
and classify such a quotient.  It is believed that (non-)abelian statistics of
topological excitations are fully described by unitary modular tensor
categories (UMTC),\cite{BK01,K062,Wang10} a notion of which is equivalent to that
of a non-degenerate unitary braided fusion category,\cite{K062,DGNO09}
abbreviated as a non-degenerate UBFC (for an introduction to category and UMTC,
see Appendix \ref{sec:cat-view-I}). 

Thus, we can use the classification of non-degenerate UBFC's\cite{RSW0777} to
classify 2+1D bosonic topological orders (see Remark\,\ref{rem:anomaly-free}) up to
invertible topological orders. In a recent paper,\cite{W150605768} we have used
such an approach to create a full list of simple  2+1D bosonic topological
orders (up to invertible topological orders).  The invertible topological
orders can be easily included by stacking with a number of layers of $E_8$
bosonic quantum Hall states.

In this paper, we develop a theory for 2+1D fermionic topological
orders without symmetry:\cite{GWW1017,GWW1332,KTT1429} 
\begin{quote}
\emph{Up to invertible topological orders, {\rm 2+1D} fermionic topological
orders without symmetry are classified by non-degenerate {\rm UBFC}'s over the
symmetric fusion category {\rm (SFC)} $\cF_0$, where the {\rm SFC} $\cF_0$ describes
a fermionic product state without symmetry, and the non-degenerate {\rm UBFC}'s
have modular extensions. 
}
\end{quote}

Several new concepts are used in the above statement.  We define SFC in
Sec.\,\ref{symmUBFC} and explain in Sec.\,\ref{symmUBFCA} why a SFC $\cE$
describes a fermionic/bosonic product state.  We explain the notion of
modular extension in Sec.\,\ref{mod-ext} and Sec.\,\ref{Bext}.

Here we briefly discuss fermionic invertible topological orders. It is believed
that all 2+1D fermionic invertible topological orders,\cite{KTT1429} also form
an abelian group $\Z$ under the stacking operation $\boxtimes$, which is
generated by the $p+\ii p$ superconductor of spinless fermions.\cite{RG0067}
The $p+\ii p$ superconductor has no non-trivial topological excitations. But
$p+\ii p$ superconductor has a non-trivial thermal Hall effect and ungappable
gapless chiral edge states with a chiral central charge $c=1/2$, and thus has a
non-trivial topological order.  The most general 2+1D fermionic invertible
topological orders can be obtained by stacking a finite number of layers of
2+1D $p\pm \ii p$ superconductors.

\def\arraystretch{1.22} \setlength\tabcolsep{3pt}
\begin{table*}[tbp] 
\caption{
A list of simple fermionic topological orders (up to invertible ones) with  $N$
types of topological excitations (including the parent fermion) and  chiral
central charge $c$ (mod 1/2). The excitations have quantum dimension $d_i$ and
spin $s_i$ (mod 1).  The searched maximum $\t N^{[i][j]}_{[k]}$ is 
1 for $N=10$.  
} 
\label{toplst10} 
\centering
\begin{tabular}{ |c|c|c|l|l|c| } 
\hline 
$N^F_c$ & $S_\text{top}$ & $D^2$ & $d_1,d_2,\cdots$ & $s_1,s_2,\cdots$ & comments \\
 \hline 
$10^F_{ 0}$ & $1.1609$ & $10$ & $1,1,1,1,1,1,1,1,1,1$ & $0, \frac{1}{2}, \frac{1}{10},-\frac{2}{5}, \frac{1}{10},-\frac{2}{5},-\frac{1}{10}, \frac{2}{5},-\frac{1}{10}, \frac{2}{5}$ & $\cF_0\boxtimes 5^B_{ 4}$\\
$10^F_{ 0}$ & $1.1609$ & $10$ & $1,1,1,1,1,1,1,1,1,1$ & $0, \frac{1}{2}, \frac{3}{10},-\frac{1}{5}, \frac{3}{10},-\frac{1}{5},-\frac{3}{10}, \frac{1}{5},-\frac{3}{10}, \frac{1}{5}$ & $\cF_0\boxtimes 5^B_{ 0}$\\
$10^F_{ 0}$ & $1.7924$ & $24$ & $1,1,1,1,\zeta_{4}^{1},\zeta_{4}^{1},\zeta_{4}^{1},\zeta_{4}^{1},2,2$ & $0, \frac{1}{2}, \frac{1}{2}, 0, \frac{1}{8},-\frac{3}{8},-\frac{3}{8}, \frac{1}{8}, \frac{1}{6},-\frac{1}{3}$ & $\cF_0\boxtimes 5^{B,b}_{-2}$\\
$10^F_{ 0}$ & $1.7924$ & $24$ & $1,1,1,1,\zeta_{4}^{1},\zeta_{4}^{1},\zeta_{4}^{1},\zeta_{4}^{1},2,2$ & $0, \frac{1}{2}, \frac{1}{2}, 0, \frac{1}{8},-\frac{3}{8},-\frac{3}{8}, \frac{1}{8},-\frac{1}{6}, \frac{1}{3}$ & $\cF_0\boxtimes 5^{B,a}_{ 2}$\\
$10^F_{ 0}$ & $1.7924$ & $24$ & $1,1,1,1,\zeta_{4}^{1},\zeta_{4}^{1},\zeta_{4}^{1},\zeta_{4}^{1},2,2$ & $0, \frac{1}{2}, \frac{1}{2}, 0,-\frac{1}{8}, \frac{3}{8}, \frac{3}{8},-\frac{1}{8}, \frac{1}{6},-\frac{1}{3}$ & $\cF_0\boxtimes 5^{B,a}_{-2}$\\
$10^F_{ 0}$ & $1.7924$ & $24$ & $1,1,1,1,\zeta_{4}^{1},\zeta_{4}^{1},\zeta_{4}^{1},\zeta_{4}^{1},2,2$ & $0, \frac{1}{2}, \frac{1}{2}, 0,-\frac{1}{8}, \frac{3}{8}, \frac{3}{8},-\frac{1}{8},-\frac{1}{6}, \frac{1}{3}$ & $\cF_0\boxtimes 5^{B,b}_{ 2}$\\
$10^F_{5/11}$ & $2.5573$ & $69.292$ & $1,1,\zeta_{9}^{1},\zeta_{9}^{1},\zeta_{9}^{2},\zeta_{9}^{2},\zeta_{9}^{3},\zeta_{9}^{3},\zeta_{9}^{4},\zeta_{9}^{4}$ & $0, \frac{1}{2}, \frac{7}{22},-\frac{2}{11},-\frac{7}{22}, \frac{2}{11},-\frac{9}{22}, \frac{1}{11}, \frac{1}{22},-\frac{5}{11}$ & $\cF_0\boxtimes 5^B_{\frc{16}{11}}$\\
$10^F_{-5/11}$ & $2.5573$ & $69.292$ & $1,1,\zeta_{9}^{1},\zeta_{9}^{1},\zeta_{9}^{2},\zeta_{9}^{2},\zeta_{9}^{3},\zeta_{9}^{3},\zeta_{9}^{4},\zeta_{9}^{4}$ & $0, \frac{1}{2},-\frac{7}{22}, \frac{2}{11}, \frac{7}{22},-\frac{2}{11}, \frac{9}{22},-\frac{1}{11},-\frac{1}{22}, \frac{5}{11}$ & $\cF_0\boxtimes 5^B_{-\frc{16}{11}}$\\
 \hline 
\end{tabular} 
\end{table*}

To develop a simple theory for 2+1D fermionic topological orders, we assume
that the  (non-)abelian statistics of topological excitations in 2+1D fermionic
topological orders is fully described by the data $(N^{ij}_k,s_i)$, where
$i,j,k$ label the types of topological excitations, $s_i$ is the spin (mod 1)
of the type-$i$ topological excitation, and $N^{ij}_k$ are the fusion
coefficients of  topological excitations.  We find the conditions
that the data $(N^{ij}_k,s_i)$ must satisfy in order to describe a 2+1D
fermionic topological order.  
By finding all the $(N^{ij}_k,s_i)$'s that
satisfy the conditions, we obtain a classification of 2+1D fermionic
topological orders (up to invertible topological orders).  If we further
include the chiral central charge $c$ of the edge states, we believe that the
data $(N^{ij}_k,s_i,c)$ describe/classify all 2+1D fermionic topological
orders (including the invertible ones).

We have numerically searched the $(N^{ij}_k,s_i,c)$ that satisfy the
conditions. This allows us to create a list of simple 2+1D fermionic
topological orders (up to invertible topological orders) (see Tables
\ref{toplst} and \ref{toplst10}).  The invertible topological orders can be
easily included by stacking with a number of layers of $p \pm \ii p$ fermionic
superconductors.

\subsection{Classify topological orders with symmetry}\label{mainproposal}

Using the fact that a bosonic/fermionic symmetry is uniquely determined by a
SFC $\cE$ (see Sec.\,\ref{sec:cat-view-II} for details),\cite{Del02} 
we propose a complete classification of 2+1D fermionic/bosonic topological
orders with symmetry: 
\begin{quote}
\emph{{\rm 2+1D} topological orders with the symmetry $\cE$ are classified by
$(\cC,\cM,c)$, where $\cC$ is a non-degenerate {\rm UBFC} over $\cE$, $\cM$ is
a modular extension of $\cC$, and $c\in \Q$ is the total chiral central
charge.}
\end{quote}
There are five main ingredients of above proposal:
\begin{enumerate}
\item By definition, UBFC describes topological excitations
and their fusion-braiding properties (\ie their non-abelian statistics).
It is clear that UBFC overlooks the edge states (\ie cannot detect invertible topological orders).

\item 
The SFC $\cE$ is a special kind of UBFC that describes the excitations in
bonson/fermion product state with symmetry. In fact, the bosonic/fermionic
symmetry is uniquely determined by $\cE$.  Thus $\cE$ is a
categorical description of symmetry.

\item
The non-abelian
    statistics of bulk topological excitations in a topological order with
    symmetry $\cE$ is described by a non-degenerate
    UBFC $\cC$ over $\cE$.
The term ``over'' in the above means: (1) the UBFC $\cC$ contains $\cE$.
In other words, $\cC$ contains all
the excitations of product state with the same symmetry. (2) the excitations in $\cE$ have trivial mutual statistics
with all the excitations in $\cC$.  The term ``non-degenerate'' means that only
the excitations in $\cE$ can have trivial mutual statistics with all the
excitations in $\cC$.

  \item 
Roughly speaking,  a modular extension corresponds to gauging all the symmetry.\cite{LG1209,CF14036491}
Up to the $E_8$ states, the edge states of a non-degenerate UBFC $\cC$
    over $\cE$, are classified by 
    the modular extensions of $\cC$ 
    (see Sec.\,\ref{mod-ext} and \ref{Bext} for detailed explanation). 
In particular, the modular extensions of $\cE$ classify  invertible topological orders with symmetry $\cE$ up to the $E_8$ states. We believe that
    they are exactly the symmetry protected topological (SPT)\cite{CLW1141,CGL1314,CGL1204} states.
  \item The remaining ambiguity, \ie the number of layers of $E_8$ states, is
    fixed by the total chiral central charge $c$.
\end{enumerate}

By combining with the theory of BF category developed in \Ref{KW1458},
the above proposal can be naturally generalize to higher dimensions, 
\begin{quote}
\emph{Up to invertible topological orders,  $(d$+$1)$D fermionic/bosonic
topological orders with/without symmetry are classified by non-degenerate
unitary braided fusion $(d-$$1)$-categories over a symmetric fusion 1-category;
the symmetric fusion 1-category, viewed as a unitary braided fusion
$(d-$$1)$-category with only trivial $k$-morphisms for $0\leq k<d$, describes a
$(d$+$1)$D fermionic/bosonic product state with/without symmetry.  We also
require that the non-degenerate unitary braided fusion $(d-$$1)$-category has a
modular extension.  } 
\end{quote}

Fermionic/bosonic topological orders with symmetry will be thoroughly studied
in an upcoming paper \Ref{KLW}. In this paper, we concentrate on 2+1D fermionic
topological orders without symmetry, which are the simplest examples of
non-degenerate UBFC's over a SFC.

\subsection{Symmetric fusion categories for bosonic/fermionic product states
with symmetry} \label{symmUBFC}

What is a SFC?  A SFC describes a bosonic/fermionic product state with/without
symmetry.  It is characterized by the quasiparticle excitations.  The SFC
that describes a fermionic product state without symmetry (denoted by $\cF_0$)
contains only two types of quasiparticles (two simple objects): the trivial
quasiparticle $\one$, and the parent fermion $f$ that forms the fermionic
system.  The SFC that describes a bosonic product state with symmetry $G$ (a
finite group) is the category of $G$-representations, denoted by $\rep(G)$ (see
Example\,\ref{expl:repG}).  The quasiparticles (the simple objects) are all
bosonic and correspond to irreducible $G$-representations.  The SFC that
describes a fermionic product state with full symmetry $G^f$, which contains,
in particular, the fermion-number-parity symmetry $\Z_2^f$,\cite{GW1441} are
the SFC of the \mbox{super-representations} of $G^f$ (see Sec.\,\ref{symmUBFCA} and
Example\,\ref{expl:srepG}), denoted by $\repgz{G^f}{}$. The quasiparticles (the
simple objects) correspond to the irreducible representations of $G^f$.
They are fermionic if $\Z_2^f$ acts non-trivially on the corresponding
representation, and bosonic if $\Z_2^f$ acts trivially.

\subsection{Relation to $G$-crossed category}\label{sec:gcross}

Note that our proposal in the bosonic cases
\begin{quote} 
\emph{{\rm 2+1D} bosonic topological orders with symmetry $G$, up to invertible
topological orders, are classified by non-degenerate {\rm UBFC}'s over
$\rep(G)$, where the non-degenerate {\rm UBFC}'s have modular extensions}.  
\end{quote}
is different, but equivalent to another proposal in \Ref{BBC1440}, using
$G$-crossed UMTC's to classify 2+1D bosonic topological orders with symmetry
$G$.
Mathematically, a non-degenerate UBFC $\cC$ over $\rep(G)$, with modular
extension $\cM$, is
related to a $G$-crossed UMTC $\cD\cong \cM_G$ via the
de-equivariantization and equivariantization processes.\cite{DGNO09}
Let $\cD_0$ be the neutral component (the full subcategory graded by the
identity element of the group $G$) of $\cD$. Note that $\cD_0$ is a UMTC with
a $G$-action. We have
\begin{align*}
  \xymatrixcolsep{7pc}
  \xymatrix{ 
    \cC\cong\cD_0^G\ar@{^{(}->}[d]_{\cC=\cen{\rep(G)}{\cM}}
    \ar@<.5ex>[r]^{\text{de-equivariantization}} 
    & \cC_G\cong\cD_0\ar@{^{(}->}[d]^{\text{neutral component}}
    \ar@<.5ex>[l]^{\text{equivariantization}} \\
    \cM\cong\cD^G\ar@<.5ex>[r]^{\text{de-equivariantization}}
    &\cM_G\cong\cD\ar@<.5ex>[l]^{\text{equivariantization}} 
  }
\end{align*}
where de-equivariantization and equivariantization are inverse to each other. This is
why we say that the two proposals are equivalent in the bosonic cases.
We will further study their relation elsewhere. 
However, our proposal has the advantage that it easily generalizes to fermionic
cases, by replacing $\rep(G)$ with $\repgz{G^f}$.

Given a symmetry $G$, not all UBFC's are over $\rep(G)$. Similarly, not all UMTC's admit a $G$-action; there are group cohomological obstructions to define
the $G$-action on a UMTC.\cite{BBC1440}
They must vanish for a consistent $G$-action on a UMTC.
However, from a non-degenerate UBFC over $\rep(G)$, there is no
further obstruction to obtain a UMTC with a $G$-action via de-equivariantization.  
\Ref{CF14036491} showed that when the obstructions do not vanish, the anomalous
symmetry action can still be realized on the surface of 3+1D systems.
To study such anomalous cases we need the higher dimensional analogs of our
proposal.

\subsection{Remarks}

\begin{rema} \label{rem:anomaly-free} {\rm
Without further announcement, all 2+1D topological orders considered in this work are anomaly-free (or closed) in the sense that they can be realized by a 2+1D lattice model with a local Hamiltonian.\cite{KW1458} 
}
\end{rema}

\begin{rema} {\rm
  We restrict ourselves to \emph{finite} symmetry groups in this work.
  The (super-)representations of finite groups form symmetric \emph{fusion}
  categories. For continuous groups, their (super-)representations still form
  symmetric tensor categories, but not \emph{fusion} categories (there are
  infinitely many different irreducible representations).
  It is not clear to what extent our results apply to cases of continuous
  groups.
}
\end{rema}

\begin{rema}  {\rm
Three types of tensor products are used in this work. We use $\boxtimes$ for the stacking product of two phases, $\otimes$ for the fusion product of particles, and $\otimes_\Cb$ for the usual tensor product of vector spaces over $\Cb$ and that of matrices with $\Cb$-entries. 
}
\end{rema}

\section{Categorical description of topological orders with symmetry} \label{sec:cat-view-II}

In this section, we give a physically motivated discussion 
on how to find a categorical description of the particle statistics in a fermionic/bosonic
topological order with symmetry. Readers who are not familiar with the categorical view of particle statistics are welcome to first read an elementary discussion of it in Appendix \ref{sec:cat-view-I}.

\subsection{Trivial topological orders with symmetry\\-- Categorical 
view of symmetry}
\label{symmUBFCA}

A 2+1D phase with trivial topological order (\ie a product state) can have
only {\it local particles}, which, by definition, are particles that can be created/annihilated by local operators.  
In a bosonic trivial phase without symmetry, there is only one type of
(indecomposable) particle: the trivial particle $\one$. When we localize the
particle by a trap, the trapped trivial particle has no internal degrees of
freedom (\ie no degeneracy) and is described by a 1-dimensional Hilbert space
$\Cb$.  For some very special traps, we may have accidental degeneracy
described a finite dimensional Hilbert space.  Such a trapped particle with
accidental degeneracy is called a composite particle, and is a direct sum of
the trivial particle.  Therefore, the bosonic product states without symmetry
can be described by the category of finite dimensional Hilbert spaces, denoted
by $\cB_0$, in which the 1-dimensional Hilbert space $\Cb$ is the trivial
particle. 

For a 2+1D product state with symmetry (given by a finite group $G$), all
the particles can be created/annihilated by local operators, and are local
excitations. They can carry additional charges from the representations of the
symmetry.
As a consequence, (indecomposable) particles in a bosonic product state with
symmetry are described by irreducible representations of $G$. Thus the trivial
topological order with symmetry is described by the category of $G$-representations, denoted by $\rep(G)$ (see also
Example\,\ref{expl:repG}).

For a fermionic product state with symmetry, we must include in $G$ the
fermion-number parity transformation $z$ ($z\neq 1$), which is involutive, \ie
$z^2=1$, and commutes with other symmetries, \ie $zg=gz$ for all $g\in G$.
Therefore, the fermonic symmetry is pair $G^f=(G,z)$. The particles in the
fermionic product state with symmetry $G^f$ still have to be classified by
irreducible representations of $G$.  However, some particles are bosonic and
some particles are fermionic: An irreducible representation is bosonic (or
fermionic) if $z$ acts as $1$ (or $-1$) in the irreducible representation.
These representations braid as bosons and fermions with trivial  mutual
statistics. Namely, by exchanging the positions of two fermions, we get an
extra $-1$ sign (see Example\,\ref{expl:srepG} for a precise mathematical
definition). 
Therefore, the particles in a fermionic product state with symmetry $G^f$ are
described by the category $\repgz{G^f}{}$, which is the same category as
$\rep(G)$ but equipped with the braidings defined according to the
fermion-number parity.  For the fermonic trivial topological order without
symmetry, there is no symmetry other than the fermion-number parity symmetry
$z$, \ie $G=\{ 1, z\}=\Z_2$ or $G^f=\Z_2^f=(\Z_2, z)$. In this case, we also
denote $\repgz{\Z_2^f}{}$ by $\cF_0$ (see also Sec. \ref{symmUBFC}).

The categories $\rep(G)$ and $\repgz{G^f}{}$ are examples of symmetric fusion
category (SFC), which is a UBFC with only trivial double braidings \ie trivial
mutual statistics (see Sec.\,\ref{ubfcsym} and Appendix \ref{mathdfn} for precise
definitions).  It turns out that all SFC's are of these types.\cite{Del02} More
precisely, an SFC $\cE$ is either $\rep(G)$ for a unique group $G$ or
$\repgz{G^f}{}$ for a unique group $G$ and a central involutive element $1\neq
z\in G$. In other words, SFC's are in one-to-one correspondence with (finite)
bosonic/fermionic symmetry groups ($G$ or $G^f=(G,z)$).
Therefore, we can refer to a given bosonic/fermionic symmetry
by a SFC $\cE$, instead of the traditional way, by groups. 
This is the categorical way to describe symmetries.

In summary, we obtain the following result. 
\begin{quote}
{\it All the excitations in a $2+$$1$D bosonic/fermionic product state with symmetry $\cE$ are local, and are described by the {\rm SFC} $\cE$. 
}
\end{quote}
Note that above statement also covers the cases without symmetry. In
particular, when $\cE=\cB_0$, it describes a bosonic trivial topological order
without symmetry; when $\cE=\cF_0$, it describes a fermonic trivial topological
order without symmetry.

\subsection{Non-trivial topological orders with symmetries} \label{sec:non-deg-ubfc}
UBFC is the natural language to describe the particle statistics (braiding and
fusion) in topological orders.
The categorical description of symmetry, using the SFC $\cE$ instead of the
symmetry group, makes it more straightforward to consider non-trivial
topological orders with symmetries.
Roughly speaking, a UBFC $\cC$ describing a non-trivial topological order with
symmetry $\cE$, must ``contain'' $\cE$ in a certain way. More precisely
\begin{enumerate}
\item $\cC$ contains local excitations carrying all the
irreducible representations of the symmetry group $G$. Mathematically, it means
that $\cC$ must contain $\cE$ (either $\rep(G)$ or $\repgz{G^f}{}$) as a full
subcategory (see Def.\,\ref{def:full-subcat}).
\item Since local excitations, by definition, can be created/annihilated by
local operators, they must have trivial mutual statistics with all particles
(including themselves). Mathematically, it means that $\cE$ lies in the
centralizer $\cen{\cC}{\cC}$ of $\cC$. The centralizer $\cen{\cC}{\cC}$ of
$\cC$ is defined as the full subcategory contains objects that have trivial
mutual braidings with all objects (including themselves). See
eqn.\,\eqref{eq:centralizer} and Def.\,\ref{def:centralizer} for precise
definitions. 
\item {\it Non-degeneracy condition}: In order for the phase to be anomaly-free (recall Remark\,\ref{rem:anomaly-free}), if a particle has trivial mutual statistics with all particles, it must be a local excitation. Mathematically, it just means that $\cen{\cC}{\cC}=\cE$. 
\end{enumerate}
A UBFC satisfying the above three properties is called a {\it non-degenerate {\rm UBFC} over $\cE$} (see also Sec.\,\ref{ubfcsym} and Def.\,\ref{def:non-deg-over-E}).\cite{DGNO09,DNO11} The precise requirements of the non-degeneracy condition on the $S$-matrix is given in Sec.~\ref{sec:concrete-data}. Note that the simplest non-degenerate UBFC over $\cE$ is just $\cE$ itself, which is nothing but the trivial topological order with the symmetry $\cE$.

In summary, we conclude that 
\begin{quote}
The bulk topological excitations in a bosonic/fermionic topological order with symmetry $\cE$ is described by a non-degenerate UBFC over $\cE$. 
\end{quote}
We describe the notion of a (non-degenerate) UBFC over $\cE$ by concrete
computable data in Sec.\,\ref{sec:concrete-data}. For precise mathematical
definition see Appendix \ref{mathdfn} or see \Ref{DGNO09,DNO11}.

In Appendix \ref{sec:loa}, we provide yet another explanation of the above proposal from the point of view of local operator algebras that define the topological excitations in a topological phase with symmetry.

\subsection{How to measure edge states categorically?}\label{mod-ext}

We have explained why a bosonic/fermionic topological order with a given symmetry $\cE$ can be naturally described by a non-degenerate UBFC $\cC$ over $\cE$. But it also raises a few puzzles. 
\begin{enumerate}
\item The particles in $\cC$ can be detected or distinguished via braiding only up to those
local excitations $\cE$. This ambiguity is protected by the symmetry. It raises
a question: how to measure $\cC$ and the symmetry $\cE$ categorically? 
\item The category $\cC$ only contains the information of the excitations in
the bulk. It does not contain enough information of the edge states. It does not describe invertible topological orders. Unlike the
no-symmetry cases, in which one can compute the central charge (mod 8) of a
UMTC to get the information of the edge states, the notion of central charge is
not defined for a non-degenerate UBFC over $\cE$. It raises a question: how to
measure the edge states of $\cC$ (or invertible topological order) categorically? 
\end{enumerate}

Since the only categorical tool is the mutual braidings, the only thing we can
do is to gauge the symmetry\cite{LG1209,CF14036491,BBC1440} by adding external particles
to the system such that newly added particles can detect old particles in
$\cE$. Clearly, there are too many ways to add external
particles. We impose the following two natural principles to the categorical
detectors: 
\begin{enumerate}
\item {\it the Principle of Efficiency}: A newly added particle should have non-trivial double braidings to at least one object in $\cE$. 
\item {\it the Principle of Completeness}: The set of all new and old particles should be able to detect each other via double braidings. In other words, they must form a bosonic anomaly-free 2+1D topological order (without symmetry). 
\end{enumerate}
In other words, a categorical measurement must be ``efficient" and ``complete".
These two principles lead us to the following precise definition of a
categorical measurement, or a modular extension of $\cC$.
\begin{quote}
A {\it categorical measurement} or a {\it modular extension} of a
non-degenerate UBFC $\cC$ over $\cE$ is a UMTC $\cM$, such that $\cC$ is a full
subcategory of $\cM$, and the only particles in $\cM$ that have trivial mutual
braidings with all particles in $\cE$ are those in $\cC$. Mathematically, it
means that the centralizer of $\cE$ in $\cM$ coincides with $\cC$, \ie
$\cen{\cE}{\cM}=\cC$ (see Def.\,\ref{def:centralizer}, Def.\,\ref{modext}).
\end{quote}

Physical realities lie in how $\cC$ can be measured or detected by other nice
categories, which, in the case, are non-degenerate UBFC's (or UMTC's).
Therefore, it is natural to require that a modular extension of a
non-degenerate UBFC over
$\cE$ always exists (see Condition\,7 in Sec.\,\ref{UBFCcnd}).
(Probably this is always true, due to Ocneanu and M\"{u}ger. See Conjecture 5.2 and Remark 5.3 in
Ref.~\onlinecite{Mue03}.) In other words, it is always possible to gauge the
symmetry $\cE$ to obtain a modular extension of a non-degenerate UBFC over
$\cE$.

When $\cE=\rep(G)$, the modular extensions of $\rep(G)$ are given by the
Drinfeld centers of a fusion category $\vect_G^\omega$ for $\omega\in H^3(G,
U(1))$,\cite{dgno2007} where $\vect_G^\omega$ is the category of $G$-graded
vector spaces twisted by $\omega$. In these cases, we see that the modular
extensions of $\rep(G)$ are consistent with the well-known
classification of SPT phases by group cohomology.\cite{CLW1141,CGL1314,CGL1204} We give more details
of this case in \Ref{KLW}.  In Sec.\,\ref{sec:fermion-bext}, we further confirm
this picture by explicitly identifying the modular extensions of $\cF_0$ with
the invertible fermionic topological orders generated by $p+\ii p$
superconductors.

Given these evidences, we believe that the modular extension is the proper
categorical way to measure the  edge states and invertible topological orders
that are missing from the categorical description of the non-degenerate UBFC's
over $\cE$. Since UMTC's fix the central charge modulo 8, the only ambiguity
left is that of $E_8$ states. This leads to our main proposal in
Sec.\,\ref{mainproposal}.

\section{Non-degenerate UBFC over a SFC} \label{sec:concrete-data}

In this section, we transform abstract data and axioms of a non-degenerate UBFC over a SFC to concrete data and equations. 

\smallskip
Due to the complexity of the axioms and the extra gauge degrees of freedom, expressing the data of a UBFC as concrete tensor entries is quite impractical. To avoid such complexity, we would like to work with the universal gauge-invariant data of a
UBFC. Similar to the eigenvalues for matrices, the characters for group
representations, for a UBFC, the gauge-invariant data are the fusion rules
$N_k^{ij}$ and the topological spins $\theta_i=\ee^{2\pi \ii s_i}$. Other
gauge-invariant data, such as quantum dimensions and $S,T$-matrices, can be
expressed in terms of $N_k^{ij}$ and $\theta_i$. These gauge-invariant 
data must satisfy finitely many algebraic equations according
to the axioms of a UBFC.  

For a UBFC over a SFC, we have a similar set of gauge-invariant data
satisfying finitely many algebraic equations. This allows us to perform a
finite search (for fixed rank) for topological orders with symmetry. 
In particular, when we choose the SFC to be $\cF_0$, this leads to
a classification and a table of simple 2+1D fermionic topological orders (see Table\,\ref{toplst} and \ref{toplst10}).

\subsection{A simple definition of a braided fusion category} \label{UBFCcnd}

A unitary braided fusion category (UBFC) (also called a unitary pre-modular
category or a unitary ribbon fusion category)
is a theory of the fusion-braiding properties of systems of anyons without the
assumption of the non-degeneracy of the mutual braidings. Examples of such
anyonic systems are those consisting of fermions, or bosons with some symmetries, as building blocks. The
building blocks (the parent bosons/fermions) have trivial mutual braiding but
can still be distinguished by fermion-number parity or other symmetry charges.
This leads to degenerate mutual braidings.

In our simplified theory, a UBFC is described by an integer tensor
$N^{ij}_k$ and a mod-1 real vector $s_i$, where $i,j,k$ run from 1 to $N$ and
$N$ is called the rank of UBFC.  We may simply denote a UBFC (the collection of
data ($N^{ij}_k,s_i$)) by $\cC$, a particle $i$ in $\cC$ by $i\in\cC$.
Sometimes it is more convenient to use abstract labels rather than 1 to $N$; we
may also abuse $\cC$ as the set of labels (particles).

Not all $(N^{ij}_k, s_i)$ describe valid UBFC.  In order to describe a UBFC,
$(N^{ij}_k,\theta_i=\ee^{2\pi \ii s_i})$ must satisfy the following
conditions:\cite{W8951,GKh9410089,BK01,RSW0777,Wang10}
\begin{enumerate} 
\item
$N^{ij}_k$ are non-negative integers that satisfy 
\begin{align} 
\label{Ncnd} &
N^{ij}_k=N_k^{ji}, \ \ N_j^{1i}=\del_{ij}, \ \ \sum_{k=1}^N N_1^{i k}N_1^{
kj}=\del_{ij},
\nonumber\\
 & \sum_{m} N_m^{ij}N_l^{m k} = \sum_{n}
 N_l^{in}N_n^{j k} \text{ or } \sum_m N^{ij}_m N_m = N_i N_ j 
\end{align} 
where the matrix $N_i$ is given by $(N_{i})_{ kj} =
N^{ij}_k$, and the indices $i,j,k$ run from 1 to $N$.  In fact $ N_1^{ij}$ defines a charge conjugation $i\to \bar i$:
\begin{align} 
N_1^{ij}=\del_{\bar ij}.  
\end{align}

\item {(\it Vafa's theorem)}\cite{V8821,AM8841,BK01,Em0207007}
$N^{ij}_k$ and $s_i$ satisfy
\begin{align}
\label{Vcnd}
\sum_r V_{ijkl}^r s_r =0 \text{ mod }1
\end{align} 
where
\begin{align}
\ \ \ \ \ \ 
V_{ijkl}^r &=  
N^{ij}_r N^{kl}_{\bar r}+
N^{il}_r N^{jk}_{\bar r}+
N^{ik}_r N^{jl}_{\bar r}
\nonumber\\
&\ \ \ \
- ( \del_{ir}+ \del_{jr}+ \del_{kr}+ \del_{lr}) \sum_m N^{ij}_m N^{kl}_{\bar m}
\end{align}

\item Quantum dimension $d_i$ is the largest
eigenvalue of the matrix $N_i$. 
Total quantum dimension ${D=\sqrt{\sum_i d_i^2}}$.
\item Topological $S,T$-matrices [see eqn. (223) in \Ref{K062}]
\begin{align} 
\label{SNsss}
S_{ij}&=\frac{1}{D}\sum_k N^{i j}_k \frac{\theta_i\theta_j}{\theta_k} d_k ,
\\
T_{ij}&=\delta_{ij}\theta_i.
\end{align} 
It is obvious that $S$ is symmetric $S_{ij}=S_{ji}$.

Under charge conjugation, 
\begin{align}
  S_{ij}=S_{i\bar j}^*,\ \theta_i=\theta_{\bar i}, \text{ or } S=S^\dag C,\ \ T=TC,
\end{align}
where the charge conjugation matrix is $C_{ij}=N_1^{ij}=\delta_{i\bar j}$.
\item {(\it Weak modularity)} Define $\Theta={D}^{-1}\sum_{i}\theta_i
  d_i^2$. [see eqn. (232) in \Ref{K062}]
  \begin{align}
    S^\dag T S=\Theta T^\dag S^\dag T^\dag.
  \end{align}

\item {\it Verlinde fusion characters}\cite{V8860}
\begin{align}
\label{Ver}
  \frac{ S_{il} S_{jl}}{ S_{1l} } =\sum_k N^{ij}_k S_{kl}.
\end{align} 
which means that, for fixed $i$, $\dfrac{S_{il}}{S_{1l}}$ is the $l^\text{th}$
eigenvalue of the fusion matrices $N_i$ with left-eigenvector $v^l$, where
$(v^l)_k=S_{kl}$ that is independent of $i$.
 
\item
There exists a UMTC, such that it contains a full sub-UBFC 
described by the data $(N^{ij}_k,s_i)$.
For details, see Sec.\,\ref{Bext}, Def.\,\ref{modext}.

\end{enumerate} 
The above conditions are necessary and sufficient (due to the condition 7) for
$(N^{ij}_k,s_i)$ to describe a UBFC.

In our simplified theory, we have assumed that a UBFC can be fully characterized
by the data $(N^{ij}_k,s_i)$.  It is not clear if such an assumption is correct
or not. So far we do not know any counter examples.  If such an assumption is
correct, then each $(N^{ij}_k,s_i)$ describes a single UBFC (\ie a single
fermionic topological order up to invertible ones).  If such an assumption is
incorrect, then above conditions are necessary and sufficient for
$(N^{ij}_k,s_i)$ to describe a quotient of UBFC.  \ie each $(N^{ij}_k,s_i)$ may
correspond to several UBFC's. 

\subsection{Non-degenerate UBFC's over a SFC and classification of
2+1D bosonic/fermionic topological orders with/without symmetry} 
\label{ubfcsym}

Two anyons $i,j$ are said to be mutually local if and only if
$S_{ij}=d_id_j/D$.  In other words, the mutual braiding (also called the double
braiding) of $i,j$ is trivial. Given a UBFC $\cC$, the local subset (or \emph{centralizer}) $\cen{\cC}{\cC}$
is the anyons that are mutually local to all anyons,
\begin{align}  \label{eq:centralizer}
  \cen{\cC}{\cC}=\{i\ |\ S_{ij}=d_id_j/D, \forall j\in\cC\}.
\end{align}

We have the following key definitions:
\begin{enumerate} \label{def:nondeg-sym}
  \item A UBFC is {\it non-degenerate} (\ie a UMTC) if ${\cen{\cC}{\cC}=\{1\}}$.
In this case
the data $(N^{ij}_k,s_i)$ satisfy additional conditions:
\begin{enumerate}
  \item $S$ is a unitary matrix.
  \item $\Theta=\exp(2\pi \ii\frac{c}{8})$, where $c$ is the chiral central
    charge.
\item Let 
\begin{equation}
 \label{nuga}
 \nu_i=\frac{1}{D^2} \sum_{jk} N_i^{jk}d_{j}d_{k}\ee^{\ii 4\pi(s_{j}-s_{k})}, 
\end{equation}
then\cite{RSW0777,Wang10} $\nu_i=0$ if $i\neq \bar i$, and $\nu_i=\pm 1$ if $i
= \bar i$.     
\end{enumerate}
The above three conditions on $(N^{ij}_k,s_i,c)$ plus those conditions in 
Sec.\,\ref{UBFCcnd} gives us a simplified theory of UMTC. 
Finding $(N^{ij}_k,s_i,c)$ satisfying those conditions allows us to produce a list of simple 2+1D bosonic topological orders.\cite{W150605768}
  \item A UBFC
    $\cE$ is {\it symmetric} (\ie a SFC) if $\cen{\cE}{\cE}=\cE$.
  \item A UBFC $\cC$
    is said to be over a SFC $\cE$ 
    if $\cE\subset\cen{\cC}{\cC} \subset \cC$.
    (Note: here we really mean that the topological data $N^{ij}_k,\theta_i,
    S_{ij},\dots$ of $\cE$ embeds as a subset of data of $\cC$.)
  \item A UBFC $\cC$ over
    $\cE$ is {\it non-degenerate} if $\cen{\cC}{\cC}=\cE$.  
\end{enumerate}
One can also find more abstract definitions of the above notions in Appendix \ref{mathdfn}.

Non-degenerate UBFC's over a SFC $\cE$ classify all 
2+1D bosonic/fermionic topological orders with/without symmetry (up to
invertible ones):
\begin{enumerate}
\item
If we choose $\cE$ to be trivial, \ie $\cE=\cB_0$, then non-degenerate UBFC's over $\cE$ 
become UMTC's, which classify all 2+1D bosonic topological orders without symmetry.
\item
If we choose $\cE$ to be the SFC for fermions, \ie $\cE=\cF_0$, then the non-degenerate UBFC's over $\cE$ classify all 2+1D fermionic topological orders without symmetry.
\item
If we choose $\cE$ to be the SFC of the representations of a group
$G$, \ie $\cE=\rep(G)$, then the non-degenerate UBFC's over $\cE$ classify all 2+1D bosonic topological orders with symmetry $G$.
\item
If we choose $\cE$ to be the SFC of the super-representations of
fermionic symmetry $G^f$ (recall Sec.\,\ref{symmUBFCA}), \ie $\cE=\repgz{G^f}{}$, then non-degenerate UBFC's over $\cE$ classify all 2+1D fermionic topological orders with fermonic symmetry $G^f$.
\end{enumerate}
The first  case has been studied in \Ref{W150605768}.  In this paper, we concentrate on the second case.  We leave the other two cases to \Ref{KLW}.

\subsection{Symmetric fusion category $\cF_0 $ for fermions}

We have proposed that non-degenerate UBFC's over $\cF_0$ to classify all
2+1D fermionic topological orders without symmetry, and the SFC $\cF_0$ gives a fermionic system without topological order. But what is $\cF_0$ in gauge invariant data?  
Let us list the topological data of $\cF_0$:
\begin{enumerate}
  \item The set of objects (particles) $\cF_0 =\{1,f\}$.
  \item The fusion coefficients $N^{ij}_k$: $N_1^{11}=N_1^{ff}=N_f^{1f}=N_f^{f1}=1$. Other entries of $N_k^{ij}$ are 0.
In other words, the particle $f$ only has a $\Z_2$ conservation:
$f\otimes f =1$ and $f\otimes 1 =f$. 
  \item 
$(\th_1,\th_f)=(1,-1)$ [\ie
$(s_1,s_f)=(0,\frac12)$ or $T_{\cF_0 }=
     \begin{pmatrix}
       1&0\\0&-1
     \end{pmatrix}$].
In other words, the particle $f$ has Fermi statistics.
  \item $S_{\cF_0 }=\frac{1}{\sqrt{2}}
     \begin{pmatrix}
       1&1\\1&1
     \end{pmatrix}$.
All the particles have trivial mutual statistics between them.
\end{enumerate}
The above data, $N^{ij}_k$ and $(s_1,s_f)=(0,\frac12)$, describes the SFC for fermions.  There is only one such  SFC for fermions.

\section{Fermionic topological orders: non-degenerate UBFC over ${\cF_0 }$}

\subsection{Conditions on $(N^{ij}_k, s_i)$ for fermionic topological orders}
\label{NsCnd}

Now we are ready to apply the general properties in Sec.\,\ref{UBFCcnd} for
a UBFC, to obtain special properties of a non-degenerate UBFC over
${\cF_0 }$.  We find that a non-degenerate UBFC over ${\cF_0 }$ (\ie 2+1D fermionic
topological orders) is described by $(N^{ij}_k,s_i)$ that satisfy the
conditions in Sec.\,\ref{UBFCcnd} plus the following conditions:
\begin{enumerate}
  \item Since $f$ is abelian, we know that for each $i$ there is a unique $j$
    such that $N_j^{fi}=1$, and for $j'\neq j$, $N_{j'}^{fi}=0$. We denote such
    $j$ by $i^f$.
    Thus, fusion with $f$ defines an involution, denoted by $i\mapsto i^f$.
    We have $(i^f)^f=i$, $N_{j}^{fi}=\delta_{ii^f}$. Also $d_{i^f}=d_i$.
  \item $f$ is mutually local to all anyons:
    \begin{align}
      S_{if}=\frac{1}{D}
      \frac{\theta_i\theta_f}{\theta_{i^f}}d_{i^f}=\frac{d_i}{D}.
    \end{align}
    Thus, we have $\theta_{i^f}=-\theta_i$.
    This also means that $i^f\neq i$ and $i^f \neq \bar i$.
  \item $N_k^{ij}$ and $S_{ij}$ has some symmetries under $i\mapsto i^f$:
\begin{align}
\label{fcnd}
N_{k}^{ij}&=N_{k}^{i^fj^f}=N_{k^f}^{i^fj}=N_{k^f}^{ij^f} ,
\nonumber\\
    S_{ij}&=S_{i j^f}.
\end{align}
    This means that if we arrange the order of labels well, the $S,T$ matrices
    have the form ${S=\tilde S\otimes_\Cb S_{\cF_0 }}$, ${T=\tilde T\otimes_\Cb T_{\cF_0 }}$.
    We may introduce the equivalence relation $i\sim i^f$.
    $\tilde S$ is indexed by the equivalent classes
    ${[i]=[i^f]}$. We shall call such equivalent classes $[i]$
    \emph{up-to-fermion}  types.
    \item Using the fact that $\cen{\cC}{\cC}=\{1,f\}$, one can show that $\tilde S$
    must be unitary. Then for the fusion of equivalent classes we have the usual
    Verlinde formula
    \begin{align}
\label{Verb}
      \t N_{[k]}^{[i][j]} & \overset{\text{def}}{=} N_k^{ij}+N_{k^f}^{ij}=\sum_{[l]}
      \frac{\tilde S_{[i][l]}\tilde S_{[j][l]}\tilde S_{[k][l]}^*}{\tilde
	S_{[1][l]}},
\nonumber\\
\t S_{[i][j]} & \text{ is symmetric and unitary}.
    \end{align}
\end{enumerate}

The above conditions plus those conditions in Sec.\,\ref{UBFCcnd} on
$(N^{ij}_k, s_i)$ give us a practical definition of non-degenerate UBFC over ${\cF_0 }$,
which classify 2+1D fermionic topological orders.

\subsection{Numerical solutions for $(N^{ij}_k, s_i)$}
\label{num}

To find $(N^{ij}_k, s_i)$'s that satisfy the above conditions plus those
conditions in Sec.\,\ref{UBFCcnd}, we may start with $(\t N_{[k]}^{[i][j]},
\tilde S_{[i][j]})$ that satisfy 
\begin{align} 
\label{Ncnd1} &
\t N^{[i][j]}_{[k]}=N_{[k]}^{[j][i]}, \ \ \t N_{[j]}^{[1][i]}=\del_{[i][j]}, 
\ \ \sum_{[k]} \t N_{[1]}^{[i][k]}\t N_{[1]}^{[k][j]}=\del_{[i][j]},
\nonumber\\
 & \sum_{[m]} \t N_{[m]}^{[i][j]}\t N_{[l]}^{[m][k]} = \sum_{[n]}
 \t N_{[l]}^{[i][n]}\t N_{[n]}^{[j][k]} 
\end{align} 
and \eqn{Verb}. We then split the value
$\t N_{[k]}^{[i][j]}$ into two parts and construct $N^{ij}_k$ via
\begin{align}
\t  N_{[k]}^{[i][j]} &= N_k^{ij}+N_{k^f}^{ij} ,\nonumber\\
N_{k}^{ij}&=N_{k}^{i^fj^f}=N_{k^f}^{i^fj}=N_{k^f}^{ij^f} ,\nonumber\\
N_{k^f}^{ij}&=N_{k^f}^{i^fj^f}=N_{k}^{i^fj}=N_{k}^{ij^f}.
\end{align}
Such $N^{ij}_k$ automatically satisfy \eqn{Ver} for a $S$ that satisfies
\eqn{fcnd}.
So we only need to check if $N^{ij}_k$ satisfies \eqn{Ncnd}.

Using
\begin{align}
\sum_k N^{ij}_k d_k = d_i d_j, \ \ \ \ d_i=d_{i^f}=d_{[i]},
\end{align}
we find that
\begin{align}
\sum_{[k]} \t N^{[i][j]}_{[k]} d_{[k]} = d_{[i]} d_{[j]}.
\end{align}
Thus $d_{[i]}$ is also the largest eigenvalue of the matrix $\t N_{[i]}$ which
is given by $(\t N_{[i]})_{[k][j]} = N^{[i][j]}_{[k]}$.
The quantum dimensions $d_i$ are already determined by $\t N^{[i][j]}_{[k]}$.

Following \Ref{W150605768},
we numerically searched $(N^{ij}_k, s_i)$'s that satisfy the above four
conditions plus the conditions 1.--6. in Sec.\,\ref{UBFCcnd}.  We only
searched $N^{ij}_k$ with maximum $\t N^{[i][j]}_{[k]}$ up to 8 for $N=4$, 5 for
$N=6$, 4 for $N=8$, and 1 for $N=10$.  The results are summarized in Tables
\ref{toplst} and \ref{toplst10}.  The entries in red do not satisfy the
condition~7 in 
Sec.\,\ref{UBFCcnd} (requiring the existence of the modular extension; see 
Sec.\,\ref{Bext} for details), and do not correspond to any valid fermionic
topological orders.  Each of the other entries corresponds to a valid fermionic
topological order (up to invertible topological orders).  In the table, we used
the notation $N^F_c$ to denote fermionic topological orders with rank $N$ and
chiral central charge $c$ (mod 1/2). The central charge $c$ is given mod 1/2
since the minimal 2+1D invertible fermionic topological order has a central
charge $1/2$.

The topological excitations are labeled by $i=1,\cdots,N$.  Note that $i=1$
always label the trivial excitation, and $i=2$ always label the excitation that
corresponds to the parent fermion $f$.  Also $2i$ and $2i-1$ always correspond
to a pair of excitations differ by  $f$:
\begin{align}
 (2i)^f=2i-1,\ \ \ (2i-1)^f=2i. 
\end{align}

We like to remark that the rank $N$ is the number of the types of topological
excitations in the fermionic topological orders, which include the parent
fermion as a non-trivial type.  In literature, most people treat the parent
fermion as a trivial type; so, the number of types of topological excitations
usually referred in literature is, in our notion, the number of
up-to-fermion types of topological excitations, $N/2$.

In the table, we also listed the quantum dimensions $d_i$ and the spin $s_i$ of
the $i^\text{th}$-type of topological excitations.  
We note that the quantum dimensions satisfy
\begin{align}
 d_id_j=\sum_k N^{ij}_k d_k.
\end{align}
So in the table the quantum dimensions $d_i$ partially represent the fusion coefficients $N^{ij}_k$.

The total quantum dimension 
\begin{align}
 D^2=  \sum_{i=1}^N d_i^2 
\end{align}
is also listed.  Note that in literature, people usually define $D_F^2=
 \sum_{i=1}^{N/2} d_{2i}^2 $ as the total quantum dimension.
The topological entanglement entropy\cite{LW0605,KP0604} is given by
\begin{align}
S_\text{top} =\frac12 \log_2 D_F^2 = \frac12 \log_2 \frac{D^2}{2} .
\end{align}

From last column of the Tables \ref{toplst} and \ref{toplst10}, we see that most fermionic
topological orders can be viewed as a stacking of a bosonic topological order
(whose label was introduced in \Ref{W150605768}) with the trivial fermionic
topological order $\cF_0$ (the fermionic product state).  Some other fermionic
topological orders can be viewed as a stacking of a bosonic topological order
with a fermionic topological order, or as a stacking of two fermionic
topological orders.  There are also fermionic topological orders that are
primitive, \ie cannot be viewed as a stacking of two simpler non-trivial
topological orders.

The simplest primitive fermionic topological order is the $4^F_{ 1/4}$
topological order.  It is the first of a sequence of primitive fermionic
topological orders with $(N^F_c,D^2) = (4^F_{ 1/4}, 13.6568), \ (6^F_{\pm
0},44.784), \ (8^F_{\pm 1/8}, 105.096)$, \etc.  Another type of primitive
fermionic topological orders are the $(N^F_c,D^2) = (8^F_{0}, 24)$ topological
orders (there are eight of them with different spins $s_i$).  This is also the
first of a sequence of primitive fermionic topological orders.

\section{Stacking operation for topological orders}

In this section, we discuss the stacking operation in details.
In particular, we describe stacking operation in terms of
$(N^{ij}_k,s_i,c)$.

\subsection{Stacking fermionic/bosonic topological order with bosonic topological order}

Suppose that we have two UBFCs, $\cC$ and $\cD$, with particles (simple objects)
labeled by $i\in\cC,\ a\in\cD$. We can construct a new UBFC by
simply stacking $\cC$ and $\cD$, denoted by $\cC\boxtimes\cD$.  The anyon labels of
$\cC\boxtimes\cD$ are pairs $(i,a),i\in\cC,a\in\cD$, and the topological data
are given by (let $\cK=\cC\boxtimes\cD$)
\begin{gather}
  (N_{\cK})_{(k,c)}^{(i,a)(j,b)}=(N_\cC)_{k}^{ij} (N_\cD)^{ab}_{c},
\nonumber \\
  s_{(i,a)}^{\cK}=s_{i}^{\cC}+s_{a}^{\cD},\ \ \ c_{\cK}=c_{\cC}+ c_{{\cD}}
\nonumber \\
    T_{\cK}=T_\cC\otimes_\Cb  T_{\cD}, 
\nonumber \\ 
S_{\cK}=S_\cC\otimes_\Cb  S_{\cD}.
\end{gather}
The above defines the stacking operation of fermionic/bosonic topological order with bosonic topological order in terms of the topological data
$(N^{ij}_k,s_i,c)$.

\subsection{Abelian fermionic topological orders}

It is proved in Ref.~\onlinecite{DGNO09} that if a non-degenerate UBFC $\cC$ over ${\cF_0 }$ is abelian, it must be the stacking of some UMTC $\cB$ with ${\cF_0 }$,
$\cC=\cB \boxtimes {\cF_0 }$. In other words, abelian fermionic topological orders
$\cC$ can always be decomposed as bosonic topological orders $\cB$ stacking
with a layer of fermionic product state (with trivial fermionic topological
order).  However, this is not always true for non-abelian cases, for example,
the $4^F_{1/4}$ primitive fermionic topological order.

\subsection{Stacking two fermionic topological orders}

When we are considering two fermionic topological orders described by two UBFC's
over ${\cF_0 }$, we need a different notion of stacking, denoted by
$\cC\boxtimes_{{\cF_0 }}{\cD}$. The physical idea is that ${\cF_0 }\subset \cC$
and ${\cF_0 }\subset {\cD}$ are the same fermion background; we would like to
identify them.  The stacking $\boxtimes$ operation defined above gives us a UBFC
$\cC\boxtimes \cD$ which is over $\cF_0\boxtimes\cF_0$. However, the
correct stacking $\boxtimes_{\cF_0}$ operation should give us a  UBFC
$\cC\boxtimes_{\cF_0} \cD$ which is still over $\cF_0$. To achieve
this (\ie to identify the two $\cF_0$'s in $\cF_0\boxtimes\cF_0$ and reduce it to
a single $\cF_0$), we introduce the equivalent relation $(i,a)\sim (i^f,a^f)$,
and the anyon labels of $\cC\boxtimes_{\cF_0} \cD$ are the equivalent
classes $[(i,a)]$.  The topological data are given by (assume that $T_\cC=
\t T_\cC\otimes_\Cb T_{\cF_0 },S_{\cD}=\t S_{\cD}\otimes_\Cb S_{\cF_0 }$ and
let $\cK=\cC\boxtimes_{\cF_0} \cD$)
\begin{gather}
  (N_{\cK})_{[(k,c)]}^{[(i,a)][(j,b)]}=(N_\cC)_{k}^{ij}
  (N_\cD)^{ab}_{c}+(N_\cC)_{k^f}^{ij} (N_\cD)^{ab}_{c^f},
\nonumber \\
  s_{[(i,a)]}^{\cK}=s_{i}^{\cC}+ s_{a}^{\cD}=s_{i^f}^{\cC}+
  s_{a^f}^{\cD},\nonumber\\
  c_{\cK}=c_{\cC}+ c_{{\cD}},
\nonumber \\
  T_{\cK}=\t T_\cC\otimes_\Cb \t T_{\cD}\otimes_\Cb T_{\cF_0 },
\nonumber \\
  S_{\cK}= \t S_\cC\otimes_\Cb \t S_{\cD}\otimes_\Cb S_{\cF_0 }.
\end{gather}
The above defines the stacking operation of two fermionic topological orders in
terms of the topological data $(N^{ij}_k,s_i,c)$.  The stacking operation
between fermionic topological orders also make the set of fermionic topological
orders into a monoid.

\section{Modular extensions of a Fermionic Topological order}
\label{Bext}

First, note that if we have a UMTC $\cB$ that contains fermions,
${\cF_0 }=\{1,f\}\subset\cB$, it is possible to construct a non-degenerate UBFC 
$\cF$ over ${\cF_0 }$ by taking the subset of anyons in $\cB$ that
are local with respect to (centralize) ${\cF_0 }$,
\begin{align}
  \cF=\cen{(\cF_0)}{\cB}=\{i\ |\ i\in \cB,\ S_{if}=d_i/D\}.
\end{align}
Such a non-degenerate UBFC over ${\cF_0 }$ describes a fermionic topological
order $\cF$.  By definition, $\cB$ is the modular extension of the fermionic topological order $\cF$.  Physically this means that we can view the parent fermion that form the fermionic topological order $\cF$ as the fermionic quasi-particle from some bosonic topological orders $\cB$. This way, we can view every fermionic topological order as a part of a bosonic
topological order.
We consider it a physical requirement that fermionic topological orders
must have modular extensions (see Sec.\,\ref{mod-ext} and Sec.\,\ref{sec:fermion-bext}). This is nothing but Condition\,7 in Sec.\,\ref{UBFCcnd}.

Such modular extensions allow us to calculate the chiral central charge of the
fermionic topological order $\cF$.  We conjecture that the chiral central
charge $c$ of all the modular extensions $\cB$ of a given fermionic topological
order $\cF$ is the same modulo 1/2.  Such a chiral central charge $c$ mod 1/2
is the chiral central charge of the  fermionic topological order.  

How do we calculate the modular extension $\cB_{\cF}$ of a fermionic
topological order $\cF$ from the data of $\cF$?  
We note that all the anyons in $\cF$ are contained in $\cB_{\cF}$, and
$\cB_{\cF}$ may contain some extra anyons.
Assume that the anyon labels
of $\cB_{\cF}$ are $\{1,f,i,j,\dots,\x x,\x y,\dots\}$, where we use underline
to indicate the additional anyons (not in $\cF$). Let $\cN^{ij}_k$,
$\cS_{ij}$ be the fusion coefficients and the \mbox{$S$-matrix} for $\cB_{\cF}$, and
$N^{ij}_k$ be the fusion coefficients for $\cF$.
Using Verlinde formula
\begin{align}
  \frac{\cS_{f\x x}}{\cS_{1 \x x}}  \frac{\cS_{f\x x}}{\cS_{1 \x x}}= \frac{\cS_{1\x
  x}}{\cS_{1 \x x}}=1,
\end{align}
we find that  $\cS_{f\x x}=\pm \cS_{1\x x}=\pm d_{\x x}/D_{\cB_{\cF}}$. But by
definition $\x x\notin \cF$, we must have $\cS_{f\x x}=-d_{\x x}/D_{\cB_{\cF}}$.
Since $\cS$ is unitary, $0=\sum_a \cS_{1a}\cS_{fa}= \sum_{i\in \cF}
d_a^2/D^2_{\cB_{\cF}}-\sum_{\x x\notin \cF} d_{\x x}^2/D^2_{\cB_{\cF}}$, therefore
\begin{align}
  \sum_{i\in \cF} d_a^2=\sum_{\x
x\notin \cF} d_{\x x}^2.
\label{qde}
\end{align}
Thus the total quantum dimension $D_{\cF}$ of $\cF$ and the total quantum
dimension $D_{\cB_{\cF}}$ of its modular extension $\cB_{\cF}$ are directly related
\begin{align}
\label{DfDbf}
 D^2_{\cF} = \frac 12 D^2_{\cB_{\cF}}.
\end{align}
The above also constraints the maximal number of additional anyons we can have.

Next we try to determine the fusion rules involving $\x x,\x y, \dots$.
By Verlinde formula
\begin{align}
  \frac{\cS_{i 1}}{\cS_{1  1}} \frac{\cS_{\x x 1}}{\cS_{1  1}}&=\sum_{j\in\cF} \cN^{i\x
  x}_j\frac{\cS_{j 1}}{\cS_{1  1}}+\sum_{\x y\notin\cF} \cN^{i\x
  x}_{\x y}\frac{\cS_{\x y 1}}{\cS_{1  1}},\\
   \frac{\cS_{i f}}{\cS_{1  f}} \frac{\cS_{\x x f}}{\cS_{1  f}}&=\sum_{j\in\cF} \cN^{i\x
  x}_j\frac{\cS_{j f}}{\cS_{1  f}}+\sum_{\x y\notin\cF} \cN^{i\x
  x}_{\x y}\frac{\cS_{\x y f}}{\cS_{1  f}}.
\end{align}
Adding the two we have $0=\sum_{j\in\cF} \cN^{i\x x}_j d_j$, thus $\cN^{i\x
x}_j=0$. Similarly we can show $\cN^{\x x \x y}_{\x z}=0$. So the fusion
coefficients of odd numbers of $\x x,\x y,\x z,\dots$ always vanish.  

Therefore, $\cN_i$ for $i\in \cF$ is block diagonal: $(\cN_i)_{j\x x} =
(\cN_i)_{\x x j}=0$, where $i,j \in \cF$ and $\x x \notin \cF$.  In other
words, 
\begin{align}
\cN_i= N_i \oplus \check N_i, 
\end{align}
where $( N_i)_{jk}=\cN^{ij}_{k}=N^{ij}_{k}$ and $(\check N_i)_{\x x \x
y}=\cN^{i\x y}_{\x x}$, $i,j,k \in \cF$, $\x x,\x y \notin \cF$.  

If we pick a charge conjugation for the additional particles $\x x\mapsto
\x{\bar x}$, the conditions for fusion rules reduce to
\begin{align}
  \cN^{i \x x}_{\x y}=\cN^{\x x i}_{\x y}=\cN^{\x{\bar{x}} \x y}_{i}=\cN^{i \x{\bar{y}}}_{\x{\bar{x}}},\\
  \sum_{k\in\cF} N^{ij}_k \cN^{k\x x}_{\x y}= \sum_{\x z \notin\cF} \cN^{i \x z}_{\x
  x} \cN^{j \x y}_{\x z}.
  \label{extN}
\end{align}
With a choice of charge conjugation, it is enough to construct (or search for) the matrices $\check N_i$
to determine all the extended fusion rules $\cN^{ij}_k$.
Then, it is straightforward to search for the spins $s_i$ for the extend fusion rules
$\cN^{ij}_k$ to form some UMTC $\cB$ and check if $\cB$ contains $\cF$.

Besides the general condition \eqref{extN}, there are also some simple
constraints on $\check N_i$ that may speed up the numerical search.
Firstly, observe that \eqref{extN} is the same as
\begin{align}
\check N_i \check N_j = \sum_{k\in \cF} N^{ij}_k \check N_k,
\end{align}
where $i,j,k \in \cF$.
This means that $\check N_i$ satisfy the same fusion algebra as $ N_i$, and
$N^{ij}_k=\cN^{ij}_k$ is the structure constant; therefore, the eigenvalues of $\check N_i$
must be a subset of the eigenvalues of $ N_i$. 

Secondly, since $\sum_{\x y\notin\cF} \cN^{i \x x}_{\x y}d_{\x y}= d_i d_{\x x}$, by
Perron-Frobenius theorem, we know
that $d_i$ is the largest eigenvalue of $\check N_i$, with eigenvector $v,
v_{\x x}=d_{\x x}$. ($d_i$ is also the
largest absolute values of the eigenvalues of $\check N_i$.)
Note that ${\check N_{\bar i} \check N_{i}= \check N_i \check N_{\bar i},}$ ${ \check N_{\bar i}=\check N_i^\dag}$.
Thus, $d_i^2$ is the largest eigenvalue of the positive semi-definite Hermitian
matrix $\check N_{i}^\dag\check N_{i}$. For any unit vector $z$ we have $z^\dag \check N_i^\dag \check N_i z\leq d_i^2$, in particular,
\begin{align}
  (\check N_i^\dag \check N_i)_{\x x\x x}=\sum_{\x y} (\cN^{i\x x}_{\x y})^2\leq
  d_i^2.
  \label{extentry}
\end{align}
The above result is very helpful to reduce the scope of numerical search
for the solutions of the conditions.

Thirdly, since $\sum_{i\in\cF}
\cN^{i\x x}_{\x x}d_i= d_{\x x}^2$, combined with \eqref{qde} we have
\begin{align}
  \sum_{i\in \cF} d_i \tr \check N_i=\sum_{i\in \cF}  d_i^2.
  \label{exttr}
\end{align}
This puts strong constraints on the traces of the matrices $\check N_i$,
especially when $d_i, d_i^2$ are not all integers (but they are alway
algebraic numbers). For example if $d_i$ is of
the form $k+\sqrt{l}, k,l\in \Z$, \eqref{exttr} essentially splits into two
independent equations: the coefficients of $\sqrt{l}$ must be equal and the
rest part must be equal. This is the case for the red entries with rank $N=4$ in Table
\ref{toplst}. We can compute that $\tr \check N_1+\tr \check N_f=4$, thus $\tr
\check N_1\leq 4$ for those
red entries. Note that $\tr \check N_1$ is exactly the number of additional
particles. Therefore, combined with \eqref{extentry} we performed a finite
search for those red entries and confirmed that they have no modular extensions.

\section{A classification of 2+1D invertible fermionic topological orders}

\subsection{Quantization of chiral central charge $c$}

Let us first review a standard argument for the quantization of chiral central
charge $c$ (see for example \Ref{KW1458,KTT1429}).  Consider a bosonic or
fermionic system with invertible topological order.  After integrating out
all the dynamical degrees of freedom, we obtain a partition function that
may contain a gravitational Chern-Simons term
\begin{align}
Z[M^3]= \ee^{\ii \frac{2\pi c}{24} \int_{M^3} \om_{3} },
\end{align}
where $\dd \om_3=p_1$ is the first Pontryagin class.  When the tangent bundle
of $M^3$ is non-trivial, the above expression $\int_{M^3} \om_{3}$ is not well
defined.  In order to define the  gravitational Chern-Simons term for arbitrary
closed space-time manifold $M^3$, we note that the oriented cobordism group
$\Om^{SO}_3=0$, \ie all closed oriented 3-manifold $M^3$ is a boundary of a
4-manifold $M^4$: $M^3=\prt M^4$.  So, we can always define the  gravitational
Chern-Simons term as
\begin{align}
 \ee^{\ii \frac{2\pi c}{24} \int_{M^3=\prt M^4} \om_{3} }
=
\ee^{\ii \frac{2\pi c}{24} \int_{M^4} p_1 } .
\end{align}

However, the same  oriented 3-manifold $M^3$ can be the boundary of two
different 4-manifolds: $M^3=\prt M^4 =\prt \t M^4$.  In order for the above
definition to be self-consistent, we require that
\begin{align}
\ee^{\ii \frac{2\pi c}{24} \int_{M^4} p_1 } 
=
\ee^{\ii \frac{2\pi c}{24} \int_{\t M^4} p_1 } ,
\end{align}
or
\begin{align}
\label{intp1}
\ee^{\ii \frac{2\pi c}{24} \int_{M^4} p_1 }=1 
\end{align}
for any closed oriented 4-manifold $\prt M^4=\emptyset$.

We note that 
\begin{align}
 \int_{M^4} p_1 = 0 \text{ mod } 3.
\end{align}
Therefore $c$ must be quantized as
\begin{align}
 c= 0 \text{ mod } 8
\end{align}
to satisfy the condition \eqn{intp1}. This implies that the central charge for
bosonic invertible topological orders must be multiple of $8$, where $c=8$ is
realized by the $E_8$ bosonic quantum Hall state.

But for fermionic invertible topological orders,
the central charge is quantized differently. This is because
$M^4$ must have a spin structure for fermion systems.
In this case\cite{E0612}
\begin{align}
 \int_{M^4_\text{spin}} p_1 = 0 \text{ mod } 48.
\end{align}
Therefore $c$ must be quantized as
\begin{align}
 c= 0 \text{ mod } \frac 12
\end{align}
for 2+1D fermionic invertible topological orders.  $c=\frac 12$ is realized by
the $p+\ii p$ fermionic superconducting state.

\def\arraystretch{1.25} \setlength\tabcolsep{3pt}
\begin{table*}[tbp] 
\caption{The 16 modular extensions of the $4^F_{1/4}$ fermionic 
topological order (in the first row).
} 
\label{bextnF24} 
\centering
\begin{tabular}{ |c|c|c|l|l| } 
\hline 
$N^{F,B}_c$ & $S_\text{top}$ & $D^2$ & $d_1,d_2,\cdots$ & $s_1,s_2,\cdots$ \\
 \hline 
$4^F_{1/4}$ & $1.3857$ & $13.656$ & $1,1,\zeta_{6}^{2},\zeta_{6}^{2}$ & $0, \frac{1}{2}, \frac{1}{4},-\frac{1}{4}$  \\
 \hline 
 \hline 
$7^B_{9/4}$ & $2.3857$ & $27.313$ & $1,1,\zeta_{6}^{2},\zeta_{6}^{2},\zeta_{6}^{1},\zeta_{6}^{3},\zeta_{6}^{1}$ & $0, \frac{1}{2}, \frac{1}{4},-\frac{1}{4}, \frac{3}{32}, \frac{15}{32}, \frac{3}{32}$  \\
$7^B_{-1/4}$ & $2.3857$ & $27.313$ & $1,1,\zeta_{6}^{2},\zeta_{6}^{2},\zeta_{6}^{1},\zeta_{6}^{3},\zeta_{6}^{1}$ & $0, \frac{1}{2},-\frac{1}{4}, \frac{1}{4}, \frac{5}{32},-\frac{7}{32}, \frac{5}{32}$  \\
$7^B_{-15/4}$ & $2.3857$ & $27.313$ & $1,1,\zeta_{6}^{2},\zeta_{6}^{2},\zeta_{6}^{1},\zeta_{6}^{3},\zeta_{6}^{1}$ & $0, \frac{1}{2}, \frac{1}{4},-\frac{1}{4}, \frac{11}{32},-\frac{9}{32}, \frac{11}{32}$  \\
$7^B_{7/4}$ & $2.3857$ & $27.313$ & $1,1,\zeta_{6}^{2},\zeta_{6}^{2},\zeta_{6}^{1},\zeta_{6}^{3},\zeta_{6}^{1}$ & $0, \frac{1}{2},-\frac{1}{4}, \frac{1}{4}, \frac{13}{32}, \frac{1}{32}, \frac{13}{32}$  \\
$7^B_{-7/4}$ & $2.3857$ & $27.313$ & $1,1,\zeta_{6}^{2},\zeta_{6}^{2},\zeta_{6}^{1},\zeta_{6}^{3},\zeta_{6}^{1}$ & $0, \frac{1}{2}, \frac{1}{4},-\frac{1}{4},-\frac{13}{32},-\frac{1}{32},-\frac{13}{32}$  \\
$7^B_{15/4}$ & $2.3857$ & $27.313$ & $1,1,\zeta_{6}^{2},\zeta_{6}^{2},\zeta_{6}^{1},\zeta_{6}^{3},\zeta_{6}^{1}$ & $0, \frac{1}{2},-\frac{1}{4}, \frac{1}{4},-\frac{11}{32}, \frac{9}{32},-\frac{11}{32}$  \\
$7^B_{1/4}$ & $2.3857$ & $27.313$ & $1,1,\zeta_{6}^{2},\zeta_{6}^{2},\zeta_{6}^{1},\zeta_{6}^{3},\zeta_{6}^{1}$ & $0, \frac{1}{2}, \frac{1}{4},-\frac{1}{4},-\frac{5}{32}, \frac{7}{32},-\frac{5}{32}$  \\
$7^B_{-9/4}$ & $2.3857$ & $27.313$ & $1,1,\zeta_{6}^{2},\zeta_{6}^{2},\zeta_{6}^{1},\zeta_{6}^{3},\zeta_{6}^{1}$ & $0, \frac{1}{2},-\frac{1}{4}, \frac{1}{4},-\frac{3}{32},-\frac{15}{32},-\frac{3}{32}$  \\
$7^B_{-5/4}$ & $2.3857$ & $27.313$ & $1,1,\zeta_{6}^{2},\zeta_{6}^{2},\zeta_{6}^{3},\zeta_{6}^{1},\zeta_{6}^{1}$ & $0, \frac{1}{2}, \frac{1}{4},-\frac{1}{4},-\frac{11}{32}, \frac{1}{32}, \frac{1}{32}$  \\
$7^B_{13/4}$ & $2.3857$ & $27.313$ & $1,1,\zeta_{6}^{2},\zeta_{6}^{2},\zeta_{6}^{3},\zeta_{6}^{1},\zeta_{6}^{1}$ & $0, \frac{1}{2},-\frac{1}{4}, \frac{1}{4},-\frac{13}{32}, \frac{7}{32}, \frac{7}{32}$  \\
$7^B_{3/4}$ & $2.3857$ & $27.313$ & $1,1,\zeta_{6}^{2},\zeta_{6}^{2},\zeta_{6}^{3},\zeta_{6}^{1},\zeta_{6}^{1}$ & $0, \frac{1}{2}, \frac{1}{4},-\frac{1}{4},-\frac{3}{32}, \frac{9}{32}, \frac{9}{32}$  \\
$7^B_{-11/4}$ & $2.3857$ & $27.313$ & $1,1,\zeta_{6}^{2},\zeta_{6}^{2},\zeta_{6}^{3},\zeta_{6}^{1},\zeta_{6}^{1}$ & $0, \frac{1}{2},-\frac{1}{4}, \frac{1}{4},-\frac{5}{32}, \frac{15}{32}, \frac{15}{32}$  \\
$7^B_{11/4}$ & $2.3857$ & $27.313$ & $1,1,\zeta_{6}^{2},\zeta_{6}^{2},\zeta_{6}^{3},\zeta_{6}^{1},\zeta_{6}^{1}$ & $0, \frac{1}{2}, \frac{1}{4},-\frac{1}{4}, \frac{5}{32},-\frac{15}{32},-\frac{15}{32}$  \\
$7^B_{-3/4}$ & $2.3857$ & $27.313$ & $1,1,\zeta_{6}^{2},\zeta_{6}^{2},\zeta_{6}^{3},\zeta_{6}^{1},\zeta_{6}^{1}$ & $0, \frac{1}{2},-\frac{1}{4}, \frac{1}{4}, \frac{3}{32},-\frac{9}{32},-\frac{9}{32}$  \\
$7^B_{-13/4}$ & $2.3857$ & $27.313$ & $1,1,\zeta_{6}^{2},\zeta_{6}^{2},\zeta_{6}^{3},\zeta_{6}^{1},\zeta_{6}^{1}$ & $0, \frac{1}{2}, \frac{1}{4},-\frac{1}{4}, \frac{13}{32},-\frac{7}{32},-\frac{7}{32}$  \\
$7^B_{5/4}$ & $2.3857$ & $27.313$ & $1,1,\zeta_{6}^{2},\zeta_{6}^{2},\zeta_{6}^{3},\zeta_{6}^{1},\zeta_{6}^{1}$ & $0, \frac{1}{2},-\frac{1}{4}, \frac{1}{4}, \frac{11}{32},-\frac{1}{32},-\frac{1}{32}$  \\
 \hline 
\end{tabular} 
\end{table*}

\subsection{Classify 2+1D invertible fermionic topological orders via modular
extentions} \label{sec:fermion-bext}

However, for each quantized $c$, do we have only one invertible fermionic
topological order, or can we have several distinct invertible fermionic
topological orders?  The above analysis of the quantization of the central
charge $c$ cannot answer this question.  Here, we would like to propose the
following conjecture to address this issue:
\begin{quote} \label{conj:1} {\rm
Up to invertible bosonic topological orders, invertible fermionic topological orders are classified by the modular
extensions of $\cF_0$.  More precisely, Let $\rm i\cF$ be an invertible
fermionic topological order and define the equivalent relation $\sim: ({\rm
i}\cF\boxtimes E_8) \sim \rm i\cF$.  The quotient \{invertible fermionic
topological orders\}$/{\sim}$ is classified by the modular extensions of
$\cF_0$.
}
\end{quote}

The modular extensions of $\cF_0$ are given by the bosonic topological orders that (a) contain a
fermion $f$ and (b) $f$ has a non-trivial mutual statistics with all other
non-trivial topological excitations.  From \eqn{DfDbf}, we see that a modular
extension of  $\cF_0$ must have a total quantum dimension $D^2=4$.  We find
that the trivial fermionic topological order ${\cF_0 }$ has 16 modular
extensions: 8 Ising type UMTC
$3^B_c$ with central charge $c=\pm 1/2,\pm 3/2,\pm 5/2,\pm 7/2$, and 8 abelian
rank-4 UMTC $4^B_c$ with central charge $c=0,\pm 1,\pm 2,\pm 3,4$ (see
\Ref{W150605768}). For a detailed exposition of the mathematical structures of these 16 UMTC's see \Ref{DGNO09}. 

We conclude that, up to 
invertible bosonic topological orders,
all invertible fermionic topological orders are classified by $\Z_{16}$
generated by the $p+\ii p$ fermionic superconducting state.  This is a
generally believed result, which is one of the reasons that motivates the
above conjecture. 

For non-trivial fermionic topological orders, we further conjecture:
\begin{quote}  {\rm
The fermionic topological orders with a given set of bulk
topological excitations $\cF$ are classified by the modular extensions of $\cF$ up to
invertible bosonic topological orders. They have the same set of bulk
topological excitations $\cF$, but different edge states.
}
\end{quote}
This a special case of our general proposal mentioned in 
Sec.\,\ref{mainproposal}.

For the  fermionic topological order of the form $\cF={\cF_0 } \boxtimes \cB$
(\ie a stacking of  trivial fermionic topological order ${\cF_0 }$ and a
bosonic topological order $\cB$), it has the modular extensions (up to
invertible bosonic topological orders) given by $\cB_{\cF}=\cB_{\cF_0 }
\boxtimes \cB$, where $\cB_{\cF_0 }$ is one of the 16  modular extensions of
$\cF_0$.  They correspond to fermionic topological orders that have the same set of bulk excitations, but different edge states.  Also, the 16 modular extensions of the $4^F_{1/4}$ primitive fermionic topological order is listed in Table \ref{bextnF24}.  Again, they
correspond to fermionic topological orders that have the same set of bulk
excitations, but different edge states.

Before we end this section, we briefly remark on the relation between the modular extensions of $\cF_0$ and the Witt groups.\cite{dmno} The 16 modular extensions of $\cF_0$ does not form a group under the stacking product $\boxtimes$ because they are not invertible. But they do form a $\Z_{16}$ group if we carefully define the stacking $\boxtimes_{\cF_0}$ for modular extensions.~\cite{KLW}
Moreover, the Witt classes of these 16 modular extensions of $\cF_0$ do form a $\Z_{16}$-subgroup of the bosonic Witt group $\cW$.\cite{dmno} This subgroup is precisely the kernel of the canonical group homomorphism $\cW \to \cW_{/\cF_0}$,\cite{DGNO09,DNO11} where $\cW_{/\cF_0}$ is the Witt group for non-degenerate UBFC's over $\cF_0$. This is not an accident, it turns out that, by taking the Witt class, the set of all
modular extensions of a generic SFC $\cE$ maps onto the kernel of the
canonical group homomorphism $\cW \to \cW_{/\cE}$,\cite{KLW}  where
$\cW_{/\cE}$ is the Witt group for non-degenerate UBFC's over $\cE$. Details
will be given in \Ref{KLW}.

\section{Examples and realizations of fermionic topological orders}

\subsection{Fermionic topological orders from the $\frac{U(1)_M}{\Z_2}$-orbifold
simple-current algebra} 

\begin{table}[tb]
\caption{\label{U1/Z2rep} The irreducible  representations
$\cV^{\frac{U(1)_M}{\Z_2}}_{i}$ of $\frac{U(1)_M}{\Z_2}$-orbifold simple
current  algebra. The second column is the conformal dimensions $h_i$ of the
corresponding primary fields.  The third column is the quantum dimensions
$d_i$ of the representations.  }
\begin{center}
\begin{tabular}{|ccc|c|}
\hline
label $i$ & $h_i$ & $d_i$ &\\
\hline
 1& 0 & 1 &\\
$j$ & 1 & 1 & \\
$\phi_M^\al$ & $M/4$ & 1 & $\al=1,2$\\
$\sigma^\al$ & 1/16 & $\sqrt M$ & $\al=1,2$\\
$\tau^\al$ & 9/16 & $\sqrt M$ & $\al=1,2$\\
$\phi_\ga$ & $\ga^2/4M$ & $2$ & $\ga=1,\cdots,M-1$\\
\hline
\end{tabular}
\end{center}
\end{table}
\begin{table*}[t] 
\caption{The $S$-matrix for the $\frac{U(1)_M}{\Z_2}$-orbifold simple current
algebra with $M=$ even. Here $\ga,\la=1,\cdots, M-1$, $\al,\bt=1,2$,
 and $\si_{\al\bt}=2\del_{\al\bt}-1$.
} \label{SmatE} 
\centering
\begin{tabular}{|c|cccccc|}
\hline
$S_{ij}$ & $1$ & $j$ & $\phi_M^\al$ & $\si^\al$ & $\tau^\al$ & $\phi_\ga$ \\
\hline
$1$ & 1 & 1 & 1 & $\sqrt M$ & $\sqrt M$ & 2 \\
$j$ & 1 & 1 & 1 & $-\sqrt M$ & $-\sqrt M$ & 2 \\
$\phi_M^\bt$ & 1 &1  &1  & $\si_{\al\bt}\sqrt M$ & $\si_{\al\bt}\sqrt M$& $2(-)^{\ga}$\\
$\si^\bt$ & $\sqrt M$ & $-\sqrt M$ & $\si_{\al\bt}\sqrt M$ & $\del_{\al\bt}\sqrt {2M}$ & $-\del_{\al\bt}\sqrt {2M}$& 0 \\
$\tau^\bt$ & $\sqrt M$ & $-\sqrt M$ & $\si_{\al\bt}\sqrt M$ & $-\del_{\al\bt}\sqrt {2M}$& $\del_{\al\bt}\sqrt {2M}$ & 0 \\
$\phi_{\la}$ &  2 & 2 & $2(-)^{\la}$ & 0 & 0 & $4\cos(\pi\frac{\ga\la}{M})$\\
\hline
\end{tabular}
\end{table*} 
\begin{table*}[t] 
\caption{The $S$-matrix for the $\frac{U(1)_M}{\Z_2}$-orbifold simple current
algebra with $M=$ odd. Here $\ga,\la=1,\cdots, M-1$, $\al,\bt=1,2$, and $\si_{\al\bt}=2\del_{\al\bt}-1$.
} \label{SmatO} 
\centering
\begin{tabular}{|c|cccccc|}
\hline
$S_{ij}$ & $1$ & $j$ & $\phi_M^\al$ & $\si^\al$ & $\tau^\al$ & $\phi_\ga$ \\
\hline
$1$ & 1 & 1 & 1 & $\sqrt M$ & $\sqrt M$ & 2 \\
$j$ & 1 & 1 & 1 & $-\sqrt M$ & $-\sqrt M$ & 2 \\
$\phi_M^\bt$ & 1 &1  & $-1$  & $\ii\si_{\al\bt}\sqrt M$ & $\ii\si_{\al\bt}\sqrt M$& $2(-)^{\ga}$\\
$\si^\bt$ & $\sqrt M$ & $-\sqrt M$ & $\ii\si_{\al\bt}\sqrt M$ & $\ee^{\pi\ii\si_{\al\bt}/4}\sqrt {2M}$ & $-\ee^{\pi\ii\si_{\al\bt}/4}\sqrt {2M}$& 0 \\
$\tau^\bt$ & $\sqrt M$ & $-\sqrt M$ & $\ii\si_{\al\bt}\sqrt M$ & $-\ee^{\pi\ii\si_{\al\bt}/4}\sqrt {2M}$& $\ee^{\pi\ii\si_{\al\bt}/4}\sqrt {2M}$ & 0 \\
$\phi_{\la}$ &  2 & 2 & $2(-)^{\la}$ & 0 & 0 & $4\cos(\pi\frac{\ga\la}{M})$\\
\hline
\end{tabular}
\end{table*}

Using the correlation function of $N_p$ simple-current operators in a conformal
field theory (CFT) (or more precisely, a simple-current algebra), we can
construct an $N_p$ electron wave
function.\cite{MR9162,BW9215,WW9455,LWW1024,WW0808,WW0809,BW0932} Such an $N_p$
electron wave function describes a purely chiral fermionic topological order.
The adjoint representation generated by the simple-current  operators
corresponds to the trivial up-to-fermion type of topological excitations (see
Sec.\,\ref{NsCnd} for an explanation of up-to-fermion type of topological excitations).
While other irreducible representations of the simple-current algebra
correspond to non-trivial up-to-fermion type of topological excitations.  The number of
the up-to-fermion types of topological excitations is given by the number of the
irreducible representations of the simple-current algebra.

For example, a bosonic topological state can be constructed through
$\frac{U(1)_M}{\Z_2}$-orbifold CFT.  The $\frac{U(1)_M}{\Z_2}$-orbifold CFT
is a simple-current algebra generate by the spin-$M$ simple current
$\psi=\cos(\sqrt{2M}\phi)$ (for details, see \Ref{DVV8985}).  Since the
conformal dimension (the spin) of the  simple-current $\psi$ is an integer $M$,
$\psi$ is an bosonic operator.  The correlation of $\psi$'s gives rise to a
many-boson wave function with a bosonic topological order (for details, see
\Ref{SW15}).  

The topological excitations in such a topologically ordered state correspond to
the irreducible representations of the $\frac{U(1)_M}{\Z_2}$-orbifold
simple-current algebra, which is listed in Table \ref{U1/Z2rep}.  The spins
$s_i$ and quantum dimensions $d_i$ of those topological excitations are given
by the conformal dimensions $h_i$,  $s_i=h_i$ mod 1, and the quantum dimensions
$d_i$ of those irreducible representations.  The $S$-matrix (\ie the mutual
statistics) of those topological excitations is given in Tables \ref{SmatE} and
\ref{SmatO}.  We denote such bosonic topological order and the correspond
UMTC as $\cB_{U(1)_M/\Z_2}=\{1,j,\phi_M^\al,\phi_\ga,\tau^\al, \si^\al \}$,
where $\al=1,2$ and $\ga=1,\cdots,M-1$.

In fact, the above UMTC $\cB_{U(1)_M/\Z_2}$ with $M=6$ is a modular extension
of the $8^F_{0}$ fermionic topological order with $s_i =(0, \frac{1}{2},
\frac{1}{2}, 0, \frac{1}{6},-\frac{1}{3},-\frac{7}{16}, \frac{1}{16})$.  
From the $S$-matrix in Table \ref{SmatE}, we see that the objects/particles in ${\cF_0 }=\{1, \phi_6^1\}$, a
subset of  $\cB_{U(1)_M/\Z_2}$, are mutually local with respect to each other.
Thus the spin-$6/4$ operator $\phi_6^1$ correspond to the parent fermion $f$.
From the $S$-matrix in Table \ref{SmatE}, we also see that the topological
excitations in $\cF=\{1,\phi_6^1, \phi_6^2,j, \phi_2,\phi_4,\tau^1, \si^1 \}$,
another subset of  $\cB_{U(1)_M/\Z_2}$, are local with respect to ${\cF_0 }$.  Thus
$\cF$ is a UBFC over ${\cF_0 }$. In fact it is a non-degenerate UBFC over ${\cF_0 }$. 

The conformal dimensions and the quantum dimensions of the topological
excitations in $\cF$ are given by $h_i= (0, \frac{3}{2}, \frac{3}{2}, 1,
\frac{1}{6},\frac{2}{3},\frac{9}{16}, \frac{1}{16})$ and $d_i= (1,
1,1,1,2,2,\sqrt 6,\sqrt 6)$.  Thus $\cF$ is the non-degenerate UBFC over ${\cF_0 }$ that
describes the $8^F_{0}$ fermionic topological order with $s_i =(0, \frac{1}{2},
\frac{1}{2}, 0, \frac{1}{6},-\frac{1}{3},-\frac{7}{16}, \frac{1}{16})$ and
$d_i= (1, 1,1,1,2,2,\sqrt 6,\sqrt 6)$ (see Table \ref{toplst}).
The fusion of such a $8^F_{0}$ fermionic topological order
is given in Table \ref{U16/Z2}.

\begin{table*}[tb]
\caption{Fusion rule $j\otimes i$ for the $8^F_{0}$ fermion topological order  with 
$d_i =( 1 , 1 , 1 , 1 , 2 , 2 , \sqrt{6} , \sqrt{6})$.}
\label{U16/Z2}
\centering
\begin{tabular}{ |c|cccccccc|}
 \hline 
$d_i$ & 1 & $1$ & $1$ & $1$ & $2$ & $2$ & $\sqrt{6}$ & $\sqrt{6}$ \\
\hline
$j\backslash i$  & $\one$ & $f$ & $a$ & $a^f$ & $\al$ & $\al^f$ & $\bt$ & $\bt^f$ \\\hline
$\one$  & $\one$  & $f$  & $a$  & $a^f$  & $\al$  & $\al^f$  & $\bt$  & $\bt^f$  \\
$f$  & $f$  & $\one$  & $a^f$  & $a$  & $\al^f$  & $\al$  & $\bt^f$  & $\bt$  \\
$a$  & $a$  & $a^f$  & $\one$  & $f$  & $\al^f$  & $\al$  & $\bt$  & $\bt^f$  \\
$a^f$  & $a^f$  & $a$  & $f$  & $\one$  & $\al$  & $\al^f$  & $\bt^f$  & $\bt$  \\
$\al$  & $\al$  & $\al^f$  & $\al^f$  & $\al$  & { $\one\oplus a^f\oplus \al^f$}  & { $f\oplus a\oplus \al$}  & $\bt\oplus \bt^f$  & $\bt\oplus \bt^f$  \\
$\al^f$  & $\al^f$  & $\al$  & $\al$  & $\al^f$  & { $f\oplus a\oplus \al$}  & { $\one\oplus a^f\oplus \al^f$}  & $\bt\oplus \bt^f$  & $\bt\oplus \bt^f$  \\
$\bt$  & $\bt$  & $\bt^f$  & $\bt$  & $\bt^f$  & $\bt\oplus \bt^f$  & $\bt\oplus \bt^f$  & { $\one\oplus a\oplus \al\oplus \al^f$}  & { $f\oplus a^f\oplus \al\oplus \al^f$}  \\
$\bt^f$  & $\bt^f$  & $\bt$  & $\bt^f$  & $\bt$  & $\bt\oplus \bt^f$  & $\bt\oplus \bt^f$  & { $f\oplus a^f\oplus \al\oplus \al^f$}  & { $\one\oplus a\oplus \al\oplus \al^f$}  \\
\hline 
 \end{tabular}
\end{table*}

The above results help us to obtain the many-body wave function that realize the
$8^F_{0}$ fermionic topological order.  In fact, naively, the correlation of
the spin-$3/2$ fermionic simple-current operator $\phi_6^1$'s 
\begin{align}
 \Psi(\{z_i\}) \propto \lim_{z_\infty\to \infty} \< \hat V(z_\infty) \prod  \phi_6^1(z_i) \>
\end{align}
gives rise to a quantum-Hall many-fermion wave function $\Psi(\{z_i\})
\ee^{-\frac14 \sum |z_i|^2}$ with the above $8^F_{0}$ fermionic topological
order.  The edge excitations of such a quantum Hall state are described by the
$\frac{U(1)_M}{\Z_2}$-orbifold CFT.\cite{WWH9476,W9927,LWW1024,Wtoprev}

However, the above construction has a problem: the correlation of $\phi_6^1$
(\ie $\Psi(\{z_i\})$) has poles as $z_i\to z_j$.  But this is only a technical
problem that can be fixed as pointed out in \Ref{SW15}.  We may put the wave
function on a lattice or adding additional factors $\prod |z_i-z_j|^3$ to make
the wave function finite.  This is a realization of the  $8^F_{0}$ topological
order.

We may also introduce three complex chiral fermions $\psi_1$, $\psi_2$, and
$\psi_3$.  This allows us to construct a four-layer quantum-Hall wave function
as the following correlation in a CFT
\begin{align}
& \Psi(\{z_i,w_i,u_i,v_i\}) \propto \< \hat V(z_\infty) \prod  c_1(z_i) c_2(w_i) c_3(u_i) c_4(v_i) \> ,
\nonumber\\
& c_i=\psi_i,\ i=1,2,3,\ \ \ \ \
c_4= \psi_1 \psi_2 \psi_3 \phi_6^1.
\end{align}
In such a four-layer quantum-Hall state, the particles in the first three layers
are fermions and the particles in the fourth layer are bosons.
Such a wave function is finite, and its edge excitations are described by the
$\frac{U(1)_M}{\Z_2}\times U^3(1)$ CFT,\cite{WWH9476,W9927,LWW1024,Wtoprev}
where $U^3(1)$ CFT describes the edge excitation of $\nu=3$ integer quantum
Hall states (generated by $\psi_i,\ i=1,2,3$).  Therefore, the $8^F_{0}$
fermionic topological order described by the wave function
$\Psi(\{z_i,w_i,u_i,v_i\})\ee^{-\frac14 \sum |z_i|^2 +|w_i|^2 +|u_i|^2 +|v_i|^2
}$ only differ from the $8^F_{0}$ fermionic topological order described by
$\Psi(\{z_i\})\ee^{-\frac14 \sum |z_i|^2}$ by an invertible fermionic
topological order of the $\nu=3$ integer quantum Hall state.

The above discussion also apply to $\frac{U(1)_M}{\Z_2}$-orbifold CFT with
$M=2+4n$. When $M=2$ (\ie $n=0$), the corresponding fermionic topological order
is the $6^F_0$  topological order with
$s_i=(0,\frac12,0,\frac12,\frac1{16},-\frac7{16})$.  The case $M=6$ (\ie $n=1$)
was discussed above. The larger $n$ gives a sequence of fermionic
topological orders.  We denote those fermionic topological orders by
$\cF_{U(1)_M/\Z_2}$.  One of its modular extensions is $\cB_{U(1)_M/\Z_2}$.

We note that fermionic topological orders $\cF_{U(1)_M/\Z_2},\ M=2+4n$, always
contain a fermionic topological excitation, apart from the parent fermion.
When those fermionic topological excitations condense into invertible integer
quantum Hall states, it changes the $\cF_{U(1)_M/\Z_2}$ topological order
to some other topological order with the same quantum dimensions $d_i$ but
different spins $s_i$.  We can see those related fermionic topological orders
in Tables \ref{toplst} and \ref{toplst10}.

\begin{table}[tb]
\caption{Fusion rule $j\otimes i$ for $4^F_{\pm 1/4}$ fermionc topological
order.  $\zeta_6^2=1+\sqrt 2$.}
\label{f4F14}
\centering
\begin{tabular}{ |c|cccc|}
 \hline 
$d_i$ & 1 & $1$ & $\zeta_{6}^{2}$ & $\zeta_{6}^{2}$ \\
\hline
$j\backslash i$ & $\one$ & $f$ & $\al$ & $\al^f$ \\\hline
$\one$  & $\one$  & $f$  & $\al$  & $\al^f$  \\
$f$  & $f$  & $\one$  & $\al^f$  & $\al$  \\
$\al$  & $\al$  & $\al^f$  & $\one\oplus \al\oplus \al^f$  & $f\oplus \al\oplus \al^f$  \\
$\al^f$  & $\al^f$  & $\al$  & $f\oplus \al\oplus \al^f$  & $\one\oplus \al\oplus \al^f$  \\
\hline 
 \end{tabular}
\end{table}

\subsection{Fermionic topological orders from the $(A_1,k)$ Kac-Moody algebra}

The $(A_1,k)$ Kac-Moody algebra (\ie the $SU(2)$ level $k$ Kac-Moody algebra),
for $k\in \Z$, also gives rise to a sequence of UMTC's. The gauge-invariant data
of $(A_1,k)$ are as follows:
\begin{itemize}
  \item The set of objects (particles) labeled by $i\in \{0,1,2,\dots,k\}$.
They carry the $SU(2)$ iso-spin $S=i/2$.
The corresponding primary fields are denoted by
$V_i^m$, $m=-\frac{k}{2},-\frac{k}{2}+1,\cdots, \frac{k}{2}$.
  \item Fusion rules: $i\otimes j= |i-j|\oplus \left(|i-j|+2\right)
    \oplus\left(|i-j|+4\right)\oplus\cdots\oplus
    \min(i+j, 2k-i-j)$.
  \item Conformal dimensions $h_i=\dfrac{i(i+2)}{4(k+2)}$.
(Spins $s_i = h_i$ mod 1.)
  \item Quantum dimensions $d_i=\zeta^i_k$.
  \item Chiral central charge $c=\dfrac{3k}{k+2}$.
\end{itemize}
The above data (fusion rules and spins) describe a bosonic topological order
denoted by $\cB_{(A_1,k)}$, whose $S$-matrix can be calculated from \eqn{SNsss}.

Observe that for $k=4l+2,\ l\in\Z$, the last particle $i=4l+2$ in
$\cB_{(A_1,{4l+2})}$ is a fermion. 
The corresponding conformal field is a simple current operator.
We identity $\cF_0 =\{0, f=4l+2\}\subset \cB_{(A_1,{4l+2})}$. Then, we have a sequence of fermionic topological orders
\begin{align}
  \cF_{(A_1,4l+2)}&=\{i\in
    \cB_{(A_1,4l+2)}|S_{i,4l+2}=d_i/D\}
    \nonumber\\
    &=\{0,2,4,\dots,4l+2\}\subset
    \cB_{(A_1,4l+2)},
\end{align}
such that $\cB_{(A_1,4l+2)}$ is a modular extension of $\cF_{(A_1,4l+2)}$.  For
$l=0$, $\cF_{(A_1,2)}\cong \cF_0 $ is the trivial fermionic topological order.
The $l=1$ case has been studied in \Ref{FV13055851}.  This sequence appears in
our numerical calculations [$4^F_{1/4}, 6^F_0, 8^F_{1/8}$ in Table
\ref{toplst}]. In fact, all fermionic topological orders in this sequence are
primitive.

For $l=1$, the simple current operator carries iso-spin-3 and is given by
$V_3^m$, $m=-3,-2,\cdots,3$, with conformal dimension $h_6=\frac 32$.  To
obtain a many-body wave function that gives rise to the $4^F_{1/4}$ fermionic
topological order, we may again introduce three complex chiral fermions
$\psi_1$, $\psi_2$, and $\psi_3$.  This allows us to construct a four-layer
quantum-Hall wave function as the following correlation in a $SU(2)_6\times
U^3(1)$ Kac-Moody algebra \cite{BW9215}
\begin{align}
& \ \ \ \ \Psi(\{z_i,w_i,u_i,v_i,m_i\}) 
\nonumber\\
&\propto \< \hat V(z_\infty) \prod  c_1(z_i) c_2(w_i) c_3(u_i) c_4^{m_i}(v_i) \> ,
\nonumber\\
& c_i=\psi_i,\ i=1,2,3,\ \ \ \ \
c_4^m= \psi_1 \psi_2 \psi_3 V_6^m.
\end{align}
In such a four-layer quantum-Hall state, the particles in the first three layers
are fermions and the particles in the fourth layer are iso-spin-3 bosons.  Such
a wave function is finite, and its edge excitations are described by the
$(A_1,6)\times U^3(1)$ CFT,\cite{WWH9476,W9927,LWW1024,Wtoprev} where $U^3(1)$
CFT describes the edge excitation of $\nu=3$ integer quantum Hall states
(generated by $\psi_i,\ i=1,2,3$).  The wave function
$\Psi(\{z_i,w_i,u_i,v_i,m_i\})\ee^{-\frac14 \sum |z_i|^2 +|w_i|^2 +|u_i|^2
+|v_i|^2 }$ gives rise to the $4^F_{1/4}$ or $\cF_{(A_1,6)}$ fermionic topological order.

The $\cB_{(A_1,6)}$ is one of the modular extensions of the $4^F_{1/4}$
fermionic topological order.  Such a modular  extension is the
$N^B_c=7^B_{9/4}$ bosonic topological order in Table \ref{bextnF24}.

\section{Summary}

In this paper, we proposed that 2+1D bosonic/fermionic topological orders with
symmetry $G$ are classified, up to invertible topological orders, by
non-degenerate UBFC's over a SFC $\cE$, where $\cE$ is the category
$\text{Rep}(G)$ of $G$-representations in bosonic cases and the category
$\repgz{G^f}{}$ for fermionic cases, where $G^f=(G,z)$ and $z\in G$ is the
fermion-number parity symmetry. The case of $G=\{1\}$ (or
$G^f=(\Z_2,z)=\Z_2^f$) corresponds to the bosonic (or fermionic) case without symmetry.

We developed a simplified theory for non-degenerate UBFC over the SFC
$\cF_0=\repgz{\Z_2^f}{}$, which allows us to obtain a list of simple fermionic
topological orders with no symmetry.  We find two sequences of primitive
fermionic topological orders $\cF_{(A_1,4l+2)}$ and $\cF_{U(1)_M/\Z_2}$.

~

We would like to thank Zheng-Cheng Gu for many very helpful discussions.  This
research is supported by NSF Grant No.  DMR-1005541, and NSFC 11274192. It is
also supported by the John Templeton Foundation No. 39901. Research at
Perimeter Institute is supported by the Government of Canada through Industry
Canada and by the Province of Ontario through the Ministry of Research. 
Liang Kong is supported by the Center of Mathematical Sciences and Applications at Harvard University.

\appendix

\section{Categorical View of Particle Statistics} \label{sec:cat-view-I}

In 3+1D, particles can have two different kinds of statistics, bosonic or
fermionic. Besides, if the system has certain physical symmetry, particles also carry group representations.  The Bose/Fermi statistics and
representations of symmetry group can be unified by the single mathematical
framework \emph{symmetric categories}.

Before giving a rigorous mathematical definition, 
here we try to give a physical picture of ``categories''.  Physically, tensor category theory can be viewed as a theory that describe quasiparticle excitations in a gapped state.  The particles (point-like excitations) correspond to objects in 
category theory, and the operators or operations acting on the particles
correspond to morphisms in category theory.  Two particles that can be
connected by local operators are regarded as equivalent and correspond to two isomorphic objects in category theory.

Under such an equivalence relation, the local operators are regarded as trivial
(or null) operations, that correspond to trivial morphism.  Other operations,
such as moving one particle around another, braiding two particles, \etc,
correspond to non-trivial  morphisms.  Those operations are described by the
product of hopping operators, \ie the string operators (or Wilson loop operators).  
In other words, local operators are trivial morphisms, while string
operators can be non-trivial morphisms.  

String operators also have an equivalence relation:
Two string operators are considered equivalent if they
\begin{enumerate}
  \item have the same matrix elements among the low energy states;
  \item or differ only by local operators.
\end{enumerate}
(Those string operators are also called logic operators in topological quantum
computing.) It is the equivalent classes of string operators that correspond to
morphisms in category theory.

Besides, if there is some physical symmetry, we require the operators to
preserve the symmetry, \ie they intertwine (commute with) the symmetry
actions.  For example, two particles, carrying different irreducible
representations of the $SO(3)$ symmetry group, cannot have morphism between
them: \ie there is no symmetry preserving operations that can change one
particle into the other.  On the other hand, if one particle carry a reducible
representation of spin-1 and spin-2, and the other particle carry a reducible
representation of spin-2 and spin-3, then there is a morphism between the two
particles (objects), (\ie symmetry preserving operations may turn the first
particle into the second particle with a non-zero amplitude).  We denote the
first particle as spin-1$\oplus$spin-2 and the second particle as
spin-2$\oplus$spin-3, and the morphism as an arrow between the two particles:
\begin{align}
 (\text{spin-1}\oplus \text{spin-2}) \to
 (\text{spin-2}\oplus \text{spin-3}) .
\end{align}
In category theory, the irreducible representations, such as spin-1, correspond
to simple objects, and the reducible representations, spin-1$\oplus$spin-2,
correspond to composite objects. The composite objects are direct sums $\oplus$ of simple objects.

If we view two particles (objects) $i$ and $j$ from far away, the two
particles can be regarded as a single particle $k$. This defines a fusion
operation $\otimes$:
\begin{align}
 i\otimes j =k.
\end{align} 
If we include such a fusion operation between objects of a category, we
get a tensor category, where $\otimes$ is also called the tensor product. 

To summarize, the equivalence classes of particles form a set of objects. If we
add arrows (morphisms) between objects, we turn the set into a category. If we
further add the fusion operation, we turn the category into a tensor category.

It is the philosophy of category theory, also the physical idea of second
quantization, that we can focus on only the operators (morphisms) while treat
particles (objects) as black boxes, but still have all the information of the
system.  In other word, the particles (objects) are defined by all their
relations (morphisms) to other  particles (objects).  The morphisms correspond
to experimental observations.  The very existence of particles (objects) is 
a consequence of
those experimental observations.

Usually, when we try to understand an object, we like to divide the object into
smaller pieces (or more basic components). If we can do that, 
we gain a better understanding of the object. This is the reductionist approach.
But there is another approach. We do not think about the internal
structure of the object, and pretend the internal structure is not there (\ie
treating the object as a black box).  (Maybe the internal structure really does
not exist.) We try to understand an object through its relations (\ie
morphisms) to all objects. In fact, we use all those relations to define
the object. In other words, there are no objects, just relations. An object is uniquely determined by its relation to all objects (called Yoneda Lemma in category theory). In other words, the very existence of an object is in the form of the relations (morphisms).  This is the philosophy of category theory. We see that category theory is essentially a theory of relations. 

On the other hand, this categorical point of view is also the point of view taken by most physicists who pretend (or perhaps just get used to claim that) they are reductionists. Indeed, from a physical point of view, there is no more fundamental reality than the interrelation between particles because what can be measured in physics are not particles but their interrelations (or interactions if you like). Perhaps, a physical object only arise as an illusion of an observer after a sophisticated process of computation based on the data from interrelations. 

Now we try to introduce the operators (morphisms) in the category of particle
statistics. One of the most important examples of non-trivial operators are
those string operators (the product of local hopping operators) that generate
\emph{braidings}. Such a string operator, exchanging the positions of two
particles $a,b$ along a given path $\gamma$, corresponds to an isomorphism
$c_{a,b}: a\otimes b \to b\otimes a$.  Since local operators are quotiented
out, the braiding operator depends only on the isotopy class of the path
$\gamma$.  In 2+1D, there are two isotopy classes of paths with winding number
$\pm 1$, clockwise and counter-clockwise. They are inverse to each other.
However, in 3+1D, clockwise and counter-clockwise paths fall into the same
isotopy class; the braiding must be the inverse of itself. Such braidings are
call symmetric. (This is what the term ``symmetric'' in ``symmetric
category'' means; it refers to ``symmetric braiding'' rather than some physical
symmetry.) Therefore, in 3+1D, the braidings of identical particles can only be
either $+1$ or $-1$, corresponding to bosonic or fermionic statistics.  A
system of such particles is described by a symmetric category.  In
contrast, in 2+1D, the braidings are allowed to be more complicated, known as
anyonic or even non-abelian statistics.  Those particles are described by a
braided fusion category, which is explained later.

Other examples of topological operators are the fusion and splitting operators.
In 3+1D, they become important if we take into account the physical symmetry.
Consider two particles, carrying two irreducible representations $U,V$ of the
symmetry group. We bring them together to form a composite particle, carrying
the tensor product representation $U\otimes V$. Usually $U\otimes V$ is not
irreducible, and can fuse into to another particle carrying irreducible
representation $W$ via symmetry preserving operations. Such a process
$f:U\otimes V\to W$ is a fusion operator, corresponding to a morphism in a
category; its Hermitian conjugate $f^\dag: W\to U\otimes V$ is a splitting
operator (another morphism), corresponding to the process of splitting one
particle into two.  We need more data to describe these fusion and splitting
operators, for example, the Clebsch-Gordan coefficients for spins.
Furthermore, if more than three particles are fused, the $6j$-symbols kicks in.
They measure the difference between fusing particles in different orders.  In
3+1D, this seems just a different way to study group representations, by
focusing on how representations fuse/split rather than how the group acts.
However, the fusion and splitting operators become very rich in 2+1D.
Because anyons do not necessarily carry group representations, the fusion and
splitting operators are much more than merely the interwiners between group
representations.
This leads to rich non-abelian statistics in 2+1D.

In summary, particle statistics in 3+1D and physical symmetry are described by
symmetric fusion categories.  In 2+1D, there are new kinds of particle
statistics beyond symmetric fusion categories (\ie Bose/Fermi statistics). But
those 2+1D statistics is still not arbitrary.

First, there is a series of self-consistent conditions among the braiding,
fusion and splitting operators. These lead to the mathematical structure of a
\emph{unitary braided fusion category} (UBFC).

Secondly, we would also assume the theory to be ``complete''.
By ``complete'' we mean that ``everything can be physically measured''.  Recall
in quantum mechanics, we assume that states have inner products.
Theoretically, physical measurements are made by taking inner products.  Here
``inner products'' are ``non-degenerate'' bilinear forms. Non-degeneracy means
that if two states produce the same inner products with all other states (the
same measurement outputs), they must be the same state.  Thus, the
non-degeneracy means the theory is ``complete''.  Now, the particle statistics
are measured by the mutual braiding,
so we expect similar ``braiding non-degeneracy''.  More
precisely, the braiding measurement is performed as follows. Assume that a
particle $a$ is waiting to be measured. We first create a pair of test
particles $i$ and its antiparticle $\bar i$, then move $i$ around $a$, \ie a
double braiding, and finally annihilate $i\bar i$. The amplitude of such
process is proportional to the $(a,i)$ entry of the topological $S$-matrix,
$S_{ai}$. So the $S$-matrix is the output of the braiding measurement, and, we
should impose the non-degeneracy condition to the $S$-matrix.

For 2+1D bosonic topological orders with no symmetry, the only (topological)
measurement is the mutual braiding. Thus, different particles should be fully
distinguished by their distinct mutual braiding statistics with other
particles. If two particles have the same mutual braiding statistics with all
other particles, then the two particle must be equivalent (\ie connected by
local operators).  [This is an application of the philosophy of category
theory: an object is defined by its relations (morphisms) with all other
objects.] Indeed, a complete set of the equivalent classes of particles (\ie
the topological excitations) are described by a UBFC such that its $S$-matrix is
non-degenerate. Such a UBFC is called {\it non-degenerate}. It is equivalent to
the notion of a \emph{unitary modular tensor category} (UMTC).\cite{K062,DGNO09}
This is why we say that the topological excitations (and their non-abelian
statistics) of a 2+1D bosonic topological order are fully described by a UMTC,
or equivalently, a non-degenerate UBFC.

\section{Topological orders with symmetry from the point of view of local
operator algebras} \label{sec:loa}

In this section, we try to explain how to obtain a tensor-categorical
description of topological bulk excitations in a 2+1D topological order with symmetry from the perspective of a local operator algebra that defines these topological excitations. 

\smallskip
Let us first recall what is known in the no-symmetry cases. Consider a 2+1D bosonic topological order without symmetry that can be
realized by a lattice model.  We have a local operator algebra $A$ acting on 
the total Hilbert space $\cH_R$ associated to a disk-like region $R$.  A
topological (particle-like) excitation localized within a disk-like
region $R$ in the lattice can be defined as a subspace of $\cH_R$. Such a
topological excitation can not be created/annihilated by any local operators. As a
consequence, a topological excitation must be a module over the local operator
algebra $A$, or an $A$-module. 
This fact was fully established in Levin-Wen models that can
realize all topological orders with gappable
boundaries.\cite{KK1251,kong-icmp12,LW1384} This fact must also holds for all
topological orders because any topological order $\cC$ can be viewed as a
sub-system of a boundary-gappable topological order $\cC \boxtimes \cD$, where
$\cD$ can be chosen to be the time-reversal conjugate of $\cC$. Then the topological
excitations in $\cC$ can all be realized as modules over a local operator
algebra in a Levin-Wen model that realizes the phase $\cC \boxtimes \cD$. The
choice of $A$ is almost never unique even for a given lattice model.  The
algebra $A$ usually depends on the choice of the region $R$. But its dependence
on $R$ is not essential as it was proved in Levin-Wen models that different
local operator algebras are all Morita equivalent. \cite{kong-icmp12} In other
words, the Morita class of $A$, or equivalently, the category of $A$-modules,
denoted by $A$-Mod, is unique. We remark that this uniqueness should hold not
only for a given Levin-Wen Hamiltonian but also for a class of Hamiltonians
connected by local perturbations.

In general, $A$ is naturally equipped with a structure (somewhat equivalent to
that of an $E_2$-algebra \cite{lurie}) such that the the category $A$-Mod is a
braided monoidal category. Moreover, due to the requirement of the unitarity in
physics, we expect more structures on $A$, such as certain $\ast$-structure and
semi-simpleness, such that the category $A$-Mod is a non-degenerate UBFC (or a
UMTC). Macroscopically, the local operator algebra $A$ is not observable and not
a topological invariant either. Instead, only its Morita class (or
equivalently, the category $A$-Mod) is a macroscopic observable and a
topological invariant.

For a bosonic topological order with a symmetry group $G$, let us consider a
lattice model realizing it. In this lattice model, we still have a local
operator algebra $A$ (not respecting the symmetry) acting on 
the total Hilbert space $\cH_R$ associated to a disk-like region $R$, in
which there is a topological excitation. As in the cases without symmetry,
we do not worry about the dependence of $A$ on the region $R$. We assume that
the existence of the lattice model realizing the bosonic topological order with
symmetry $G$ is equivalent to the existence of a local operator algebra $A$
equipped with a $G$-action, \ie a group homomorphism $f: G \to \text{Aut}(A)$.
Actually, if $G$ is on-site, $G$ should also act on $\cH_R$ as local operators in $A$. As a result, there is a natural group homomorphism $G \to \text{Aut}(A)$, defined by $g \mapsto (a\mapsto gag^{-1})$ for $a\in A$ and $g\in G$. Also note that $a\mapsto gag^{-1}$ is an
algebraic isomorphism. Therefore, the microscopic data $(A,f)$ completely
determines the topological order with symmetry. What we would like to do is to
use the pair $(A,f)$ as the initial data to derive a natural macroscopic
description of this topological order with symmetry $G$. 

Note that the final macroscopic observables should respect the symmetry $G$ in
some sense.  In the microscopic world, the local operators that respect the
symmetry $G$ are those living in the fix-point algebra $A^G:=\{ a\in A|\,
ga=ag, \forall g\in G\}$, which is a sub-algebra of $A$.  Naively, it seems
that the category of $A^G$-modules, denoted by $A^G$-Mod, should be a natural
choice for the categorical description of the topological excitations in this
topological order with symmetry. But this naive choice is not good for many
reasons. The main reason is that we lose a lot of information in the process of
replacing ``$A$ with a $G$-action" by $A^G$. What we would like to do is to
find the correct replacement of the category ``$A^G$-Mod".  We do that in two
steps. In the first step, we carefully throw away the right amount microscopic
data in ``$A$ with a $G$-action" so that all the macroscopic data remain
intact; in the second step, we try to find a fix-point construction which lose
no more information. 

Similar to the no-symmetry cases, the macroscopic data of ``$A$ with a
$G$-action" is encoded in its ``Morita class". Therefore, the first step
amounts to find a proper notion of the category of modules over ``$A$ with a
$G$-action". It turns out that a $G$-action on $A$ naturally determines a
$G$-action on the category $A$-Mod as functors. More precisely, assuming that
$G$ is on-site for convenience, an $A$-module $M$, \ie a pair $(M, \rho:
A\otimes_\Cb M\to M)$, can be twisted by an element $g\in G$ to give a new
$A$-module $(M, \rho^g)$ with the action $\rho^g$ defined by
$\rho^g(a\otimes_\Cb m) = \rho(gag^{-1}\otimes_\Cb m)$. For each $g\in G$,
there is a functor $T_g: \mbox{$A$-Mod} \to \mbox{$A$-Mod}$ which maps
$(M,\rho)$ to $(M,\rho^g)$ and maps an $A$-module map $M\to N$ to the same
linear map (which automatically intertwines the actions $\rho^g$). We expect
that this functor $T_g$ also respects the monoidal and braiding structures on
$A$-Mod. Namely, $T_g$ is a braided monoidal equivalence. These functors $T_g,
\forall g\in G$, give arise to a $G$-action $T: \hat{G} \to
\text{Aut}^{br}(\mbox{$A$-Mod})$ on $A$-Mod, where $\hat{G}$ is the monoidal
category with objects given by elements in $G$ and morphisms given by identity
morphisms. Recall that only the Morita class of ``$A$ with a $G$-action"
is macroscopically meaningful.  Moreover, if we equip the category $A$-Mod with
the forgetful functor to $\mathrm{Vec}$, each $g$-action on $A$ for $g\in G$
can be recovered from the functor $T_g$ by the unique natural isomorphism from
the identity functor to $T_g$. Therefore, this ``$A$-Mod with $G$-action" can be regarded as the Morita class of ``$A$ with a $G$-action". 
This already suggests that a proper categorical
description of a topological order with symmetry is given by a non-degenerate
UBFC $\cC$ equipped with a $G$-action $T$, \ie a pair $(\cC,T)$. In particular,
the trivial phase with symmetry $G$ is given by the trivial non-degenerate UBFC
$\cB_0$ with a $G$-action. 

Note that the pair $(\cC,T)$ is not a $G$-invariant description since
the $G$-action is explicit.  So we take the second step to find a
$G$-invariant description. This can be achieved by simply replacing the
category $\cC$ with a $G$-action by the fix-point category $\cC^G$, which
consists of those objects in $\cC$ that is invariant under the $G$-action, \ie
those objects $X \in \cC$ such that $T_g(X) \simeq X, \forall g\in G$. The
category $\cC^G$ is also called the equivariantization of $(\cC,T)$ (see
\Ref{DGNO09} for a precise definition). The category $\cC^G$ turns out to be a
non-degenerate UBFC over $\rep(G)$,\cite{DGNO09} which is a UBFC with its
centralizer given by $\rep(G)$. For example, for the trivial phase
$(\cB_0,T)$ with symmetry $G$, the category $\cB_0^G$ is nothing
but $\rep(G)$. Different from the replacement of ``$A$ with a $G$-action" by
$A^G$, which losses information, that of ``$\cC$ with a $G$-action" by $\cC^G$
losses no information at all. Indeed, one can recover the former structure from
the later one by a condensation process.\cite{DGNO09} Mathematically, the
2-category of non-degenerate UBFC's equipped with a $G$-action is canonically
equivalent to that of non-degenerate UBFC's over $\rep(G)$.\cite{DGNO09}
Therefore, this notion of a non-degenerate UBFC over $\rep(G)$ is the correct
replacement to the category ``$A^G$-Mod" that we are looking for. 

In practice, working with the notion of a non-degenerate UBFC over
$\rep_G$ has more advantages than working with a non-degenerate UBFC with a
$G$-action. For example, it can be generalized easily to fermonic topological
orders with/without symmetry by replacing $\rep_G$ by a SFC $\cE$,
which determines bosonic/fermionic symmetry uniquely. An object in $\cE$ should
be viewed as a local excitation (as a trivial $A^G$-module). It can be created/annihilated by local operators that break the symmetry. It implies
that it cannot have any non-trivial double braidings with any non-trivial
topological excitations. Therefore, a SFC $\cE$ should be viewed as
the categorical description of the trivial phase with symmetry. 

In a summary, our analysis leads us to the proposal in Sec.\,\ref{sec:non-deg-ubfc} that
{\it the bulk excitations in a {\rm 2+1D} topological order with symmetry
$\cE$ are described by
  a non-degenerate {\rm UBFC} over $\cE$}.

\section{Mathematical Definitions}
\label{mathdfn}

For the reader's convenience, we collect the some relevant mathematical definitions
in this section. We would assume a basic knowledge on tensor category theory. Readers can consult with \Ref{DGNO09,DNO11} for more details. 

\begin{dfn}
A {\it fusion category} is a rigid semisimple $\C$-linear tensor category, which has only
finitely many isomorphism classes of simple objects, and has finite dimensional hom spaces, and the unit object is
simple. A {\it braided fusion category} is a fusion category endowed with a braiding satisfying the hexagon equations. (For a detailed definition, see e.g. Refs.~\onlinecite{BK01,K062}).
\end{dfn}
For physical reasons, we would assume that all the categories
are \emph{unitary}, \ie one can take the Hermitian conjugate of the morphisms
(physically they are operators between Hilbert spaces), and such Hermitian
conjugate is compatible with the fusion and braiding structures.
A unitary fusion category has a canonical spherical structure.\cite{K062} As a result, a unitary braided fusion category (UBFC) is automatically a ribbon category, or a pre-modular category.

\begin{dfn}
  The pair of objects $X,Y$ in a UBFC $\cC$ are said to \emph{centralize} (mutually local
  to) each other if
  \begin{align}
   c_{Y,X} \circ c_{X,Y}=\mathrm{id}_{X\otimes Y},
  \end{align}
  where $c_{X,Y}: X\otimes Y\xrightarrow{\simeq} Y\otimes X$ is the braiding in $\cC$.
  If $X,Y$ are simple, this is equivalent to
  \begin{align}
    S_{XY}=d_X d_Y/D,
  \end{align}
  where $S$ is the $S$-matrix.
\end{dfn}
Physically, two particles ``centralize'' or ``mutually local to'' each
other means that the two particles have the trivial mutual statistics.

\begin{dfn} \label{def:full-subcat}
A {\it full subcategory} $\cD$ of the category $\cC$ is a subcategory of $\cC$ such that every morphism in $\cC$ between two objects in $\cD$ is also a morphism in $\cD$, \ie $\hom_\cD(x,y)=\hom_\cC(x,y)$ for all $x,y\in \cD$. 
\end{dfn}

\begin{dfn}
\label{def:centralizer}
  Given a full subcategory $\cD$ of a braided fusion category $\cC$, the \emph{centralizer} of $\cD$ in $\cC$, denoted by $\cen{\cD}{\cC}$, 
  is the full subcategory of objects in $\cC$ that centralize all the objects in
  $\cD$. In particular, $\cen{\cC}{\cC}$ is called the {\it centralizer} of $\cC$. 
\end{dfn}

\begin{dfn} \label{def:sym}
A \emph{symmetric fusion category} (SFC) $\cE$ is a UBFC such that $\cen{\cE}{\cE}=\cE$. In other words,
  $\cE$ is symmetric if 
   $c_{Y,X} \circ c_{X,Y}=\mathrm{id}_{X\otimes Y}$ for objects $X,Y\in
   \cE$.
\end{dfn}

This means that all particles in a SFC have trivial mutual statistics with respect to each other.
SFC's are closely related to physical symmetries (groups).

\begin{expl} \label{expl:repG} {\rm 
For a finite group $G$, the category of $G$-representations, denoted by $\mathrm{Rep}(G)$, is an example of SFC. 
The category $\mathrm{Rep}(G)$ is equipped with the tensor product $\otimes$ given by the usual vector space tensor product $\otimes_\Cb$ and the standard symmetric braiding:
\begin{align}
  c_{X,Y}(x\otimes_\Cb y)= y\otimes_\Cb x, \quad \forall x\in X, y\in Y.
\end{align}
In particular, the category $\cB_0=\rep(\{1\})$ 
corresponds to bosonic systems without symmetry.
}
\end{expl}

\begin{expl}  \label{expl:srepG} {\rm
Let $G^f$ be a pair $(G,z)$, where $G$ is a finite group and $z$ is involutive central non-trivial element in $G$, i.e. $z^2=1$, $zg=gz$ for all $g\in G$, and $z\neq 1$. Such element $z$ acts on $G$-representations as the fermion-number parity, \ie $z x= x$ if $x$ is even and $zx=-x$ if $x$ is odd. In other words, the pair
$\Z_2^f=(\{1,z\}, z)$ is the fermion-number-parity subgroup of $G$.
The modified braiding is 
\begin{align} \label{eq:f-braid}
  c_{X,Y}(x\otimes_\Cb y)= \begin{cases}-y\otimes_\Cb x, &x,y \text{ both
    odd},\\
    y\otimes_\Cb x, &\text{otherwise}.\end{cases} 
\end{align}
The category $\repgz{G^f}{}$ is the same category as $\rep(G)$ but equipped with the modified braiding defined in eqn.\,\eqref{eq:f-braid}. $\repgz{G^f}{}$ is also an example of SFC. 
It is a ``super'' or ``fermionic'' version of
$\mathrm{Rep}(G)$, that describes fermionic symmetries.
In particular, $\repgz{\Z_2^f}{}=\cF_0$ corresponds to
fermionic systems without symmetry.
}
\end{expl}

By Deligne's theorem~\cite{Del02}, a SFC is equivalent
to either $\rep(G)$ or $\repgz{G^f}{}$.

\begin{dfn}  \label{def:non-deg-over-E} 
  A UBFC $\cC$ over a SFC $\cE$ is a UBFC $\cC$ containing $\cE$ as a full fusion subcategory in its centralizer $\cen{\cC}{\cC}$. A UBFC $\cC$ over $\cE$ is called {\it non-degenerate} if $\cE= \cen{\cC}{\cC}$.
\end{dfn}

  If we take $\cE=\cB_0$, we recover to the usual definition of unitary
  modular tensor category (UMTC).

\begin{dfn}
\label{modext}
  A {\it modular extension} of a non-degenerate UBFC $\cC$ over a SFC $\cE$ is 
  a UMTC $\cM$ such that it contains $\cC$ as a full subcategory and $\cen{\cE}{\cM}=\cC$.
\end{dfn}

Mathematically, the notion of a UBFC $\cC$, or a UBFC $\cC$
over $\cE$, is self-contained. All the definitions and conditions can be
checked within $\cC$. There is no need to require $\cC$ to be embedded into a
larger UMTC. However, in our simple definition in the main text, since only the gauge
invariant data are used, we are not sure if the data can always be concretely realized
by a UBFC. So we added the extra condition 7 in Sec.\,\ref{UBFCcnd},
which requires that the UBFC $\cC$ can be embedded into a UMTC. The physical idea is
that although the braiding in $\cC$ may be degenerate, the degeneracy can
always be resolved by adding extra particles (extending to a UMTC). 
Otherwise the braiding degeneracy cannot be measured, which is unphysical (see also Sec.\,\ref{mod-ext}.) It was conjectured that a UBFC can always be fully embedded into
a UMTC (See Conjecture 5.2 in Ref.~\onlinecite{Mue03}). 
This conjecture, if true, justifies the extra condition 7 in Sec.\,\ref{UBFCcnd}.

\bibliography{../../bib/wencross,../../bib/all,../../bib/publst,./local}
\end{document}